\documentclass[fleqn,usenatbib]{mnras}

\usepackage{color}
\usepackage[T1]{fontenc}
\usepackage{ae,aecompl}
\usepackage{xcolor}
\usepackage{ulem}
\newcommand\sbullet[1][.5]{\mathbin{\vcenter{\hbox{\scalebox{#1}{$\bullet$}}}}}

\usepackage{graphicx}	% Including figure files
\usepackage{amsmath}	% Advanced maths commands
\usepackage{cases}
\usepackage{adjustbox}

\usepackage{graphicx}	% Including figure files
\usepackage{amssymb}	% Extra maths symbols
\usepackage{newtxtext,newtxmath}
\usepackage{cases}
\usepackage{array,multirow}
\usepackage{upgreek}
\usepackage{textcomp}

\newcommand{\Msol}{\;M_{\odot}}
\newcommand{\Rsol}{\;R_{\odot}}
\newcommand{\gram}{\;\mathrm{g}}
\newcommand{\cm}{\;\mathrm{cm}}

\newcommand{\km}{\;\mathrm{km}}
\newcommand{\AU}{\;\mathrm{AU}}

\newcommand{\yr}{\;\mathrm{yr}}
\newcommand{\mstar}{\;M_{\star}}
\newcommand{\Mbh}{M_{\sbullet[0.7]}}
\newcommand{\Mbhp}{M_{\sbullet[0.7],1}}
\newcommand{\Mbhs}{M_{\sbullet[0.7],2}}

\title[MicroTDE by BBH]{Close Encounters of Stars with Stellar-mass Black Hole Binaries}

\author[T. Ryu et al.]{
Taeho Ryu$^{1,2}$,\thanks{E-mail: tryu@mpa-garching.mpg.de}
Rosalba Perna$^{3,4}$,
Yi-Han Wang$^{3}$
\\
$^{1}$ Max Planck Institute for Astrophysics, Karl-Schwarzschild-Strasse 1, 85748 Garching, Germany\\
$^{2}$ Physics and Astronomy Department, Johns Hopkins University, Baltimore, MD 21218, USA\\
$^{3}$ Department of Physics and Astronomy, Stony Brook
  University, Stony Brook, NY 11794-3800, USA\\
$^{4}$ Center for Computational Astrophysics, Flatiron Institute, New York, NY 10010, USA
}

\date{Accepted XXX. Received YYY; in original form ZZZ}

\pubyear{2022}

\begin{document}
\label{firstpage}
\pagerange{\pageref{firstpage}--\pageref{lastpage}}
\maketitle

\begin{abstract}
Many astrophysical environments, from star clusters and globular clusters to the disks of Active Galactic Nuclei, are characterized by frequent interactions between stars and the compact objects that they leave behind. Here, using a suite of $3-D$ hydrodynamics simulations, we explore the outcome of close interactions between $1M_{\odot}$ stars and  binary black holes (BBHs) in the gravitational wave regime, resulting in a tidal disruption event (TDE) or a pure scattering, focusing on the accretion rates, the back reaction on the BH binary orbital parameters and the increase in the binary BH effective spin. We find that TDEs can make a significant impact on the binary orbit, which is often different from that of pure scattering. Binaries experiencing a prograde (retrograde) TDE tend to be widened (hardened) by up to $\simeq 20\%$. Initially circular binaries become more eccentric by $\lesssim 10\%$ by a prograde or retrograde TDE, whereas the eccentricity of initially eccentric binaries increases (decreases) by a retrograde (prograde) TDE by $\lesssim 5\%$. Overall a single TDE can generally result in changes of the gravitational wave-driven merger time scale by order unity. The accretion rates of both black holes are very highly super-Eddington, showing modulations (preferentially for retrograde TDEs) on a time scale of the orbital period, which can be a characteristic feature of BBH-driven TDEs. Prograde TDEs result in the effective spin parameter $\chi$ to vary by  $\lesssim 0.02$ while $\chi\gtrsim -0.005$ for retrograde TDEs. 
\end{abstract}

\begin{keywords}
black hole physics -- gravitation -- stellar dynamics -- gravitational wave
\end{keywords}

%%%%%%%%%%%%%%%%%%%%%%%%%%%%%%%

\section{Introduction}

The last few years, especially since the discovery of the first BBH merger in gravitational waves \citep[GWs,][]{Abbott2016first}, have seen a renewed interest in stellar-mass BHs and in their dynamical interactions in dense stellar environments. In addition to the standard channel of binary star evolution (e.g. \citealt{Portegieszwart1998,Voss2003,Podsiadlowski2004,Belczynski2008,Demink2016,Mandel2016,Breivik2016,Giacobbo2018,Vignagomez2018,Mapelli2019,Broekgaarden2021,Perna2022}), dynamical formation of BBHs constitutes an important formation pathway in dense environments, such as nuclear and globular clusters (e.g. \citealt{Portegieszwart2000,Downing2010,Samsing2014,Rodriguez2015,Antonini2016,Fragione2019,Mapelli2021}) and young star clusters (e.g. \citealt{Banerjee2010, Ziosi2014,Perna2019,Rastello2019,Kumamoto2020,Dicarlo2020,Kremer2020}), as well as the disks of Active Galactic Nuclei (AGNs) (e.g. \citealt{Oleary2009,McKernan2012,Antonini2016b,Stone2017,Rasskazov2019,Tagawa2020,Wang2021AGN}). 
In these dense environments, further dynamical interactions are expected from encounters of stars with both isolated and binary BHs. These encounters can lead to a variety of outcomes, from physical 'collisions' in which the star engulfs the BH (e.g. \citealt{Fryer1998}), to 
distant encounters in which the structure of the star 
is not dramatically perturbed (e.g. \citealt{Samsing2019a}). Depending on the characteristics of  the encounter, the exchange of energy and angular momentum can lead to a dynamical reconfiguration of the system (such as for example a capture resulting in a triple), or simply a change in the orbital parameters of the BH binary (e.g. \citealt{Hamers2019,Wang2021}). 

A fraction of these close interactions between stars and BH binaries (BBHs)  will lead to 
tidal disruption events by the stellar-mass BHs, also called micro-TDEs (see e.g. \citealt{Perets2016}).
TDEs of BH binaries have been shown to play a potentially important role also in altering the spin magnitudes of the BHs after accretion of the star debris \citep{Lopez2019}.
In addition, it has  been noted how TDEs by BBHs can be used to constrain the formation history of star clusters \citep{Samsing2019}. 
Rates for TDEs by stellar BHs, whether isolated on in binaries, have been evaluated in AGN disks \citep{Yang2021} and in globular clusters \citep{Perets2016, Kremer2019}, in nuclear star clusters \citep{Fragione2021}, and in young stellar clusters \citep{Kremer2021}.
 
Despite their potential importance as discussed above, micro-TDEs, unlike TDEs by SMBHs (see e.g. \citealt{Stone2019} for a recent review) have not
received much attention, with a few exceptions
(e.g., \citealt{Perets2016,Wang2021,Kremer2022} for the single BH case, and \citealt{Lopez2019} for the BBH one). In addition to their relevance for the interpretation of LIGO/Virgo observations \citep{LIGOVIRGO}, 
studies of TDEs are especially important at this time, since
the number of detectable TDEs is expected to increase by  at least two orders of magnitude with both ongoing surveys, such as eRosita\footnote{https://erosita.mpe.mpg.de} and the Zwicky Transient Facility (ZTF)\footnote{https://www.ztf.caltech.edu}, but especially with
the upcoming Vera Rubin Observatory (VRO)\footnote{https://www.lsst.org}.

Here we perform a detailed and extensive numerical investigation of the outcome of close encounters of stars with a BBH, focusing on LIGO/Virgo BBHs (i.e. BBHs with merger times due to gravitational waves shorter than the Hubble time). We perform a wide parameter exploration, giving special emphasis both to a study of the accretion rate curves (which are relevant for the electromagnetic counterparts), as well as on the back reaction on the binary, and in particular on the effect on its semi-major axis and the eccentricity, as well as changes in the BH spins. All of these properties are relevant for the interpretation of LIGO/Virgo observations. 
We note that, for simplicity of language and notation, we will loosely use the term "TDE" for all the encounters studied here, albeit only a subset of them strictly qualifies under the standard definition, according to which the pericenter radius is larger than the radius of the star (or otherwise it would be a direct collision). 

The paper is organized as follows: \S~\ref{sec:method} discusses the ingredients of the numerical simulations, from the numerical methods to the initial conditions.  Results are reported in \S~\ref{sec:result} for 29 simulations, varying the initial orbital BBH separation and eccentricity, the phase of the binary at closest approach of the star, the mass ratio of the two BHs in the binary, and the angle of the star with respect to the orbital plane of the BBH.
We discuss their astrophysical implications such as the expected observational signatures in \S~\ref{fig:implication}, and we summarize and conclude in \S~\ref{sec:summary}.

\section{Simulation details}\label{sec:method}

\subsection{Numerical Method}

We perform a suite of $3-D$ Newtonian hydrodynamics simulations for close encounters between a main-sequence star and stellar-mass BBHs using the smoothed particle hydrodynamics (SPH) code {\tt Phantom}~\citep{Phantom}. We model the BHs using sink particles (\S\ref{subsec:binary}) and the star using smoothed particles or SPH particles (\S\ref{subsec:star}). For TDEs by SMBHs, relativistic effects can be important because the stars are disrupted at $r\lesssim 20 (\Xi/0.48) (\Mbh/10^{6}\Msol)^{-2/3}(M_{\star}/1\Msol)^{0.55} r_{\rm g}$ \citep{Ryu+2020a}, where $\Xi$ is a correction factor to the nominal tidal radius $r_{\rm t}$ to  account for realistic stellar structure and relativistic effects, and $r_{\rm g}=G\Mbh/c^{2}$ is the gravitational radius. However, for stellar-mass BHs, TDEs occur at much greater distances (e.g., $10^{4}-10^{6} r_{\rm g}$ in our simulations). Therefore, relativistic effects are expected to be negligible. 

We adopt the equation of state implemented in {\tt Phantom} which is inclusive of the radiation pressure,
\begin{align}
     P = \frac{\rho k_{\rm B}T}{\mu m_{\rm p}} + \frac{4\sigma}{3c}T^{4},
\end{align}
where $P$ is the total pressure, $\rho$ the density, $k_{\rm B}$ the Boltzmann constant, $\sigma$ the Stefan-Boltzmann constant, T the temperature and $\mu=0.62$ the mean molecular weight and $m_{\rm p}$ the proton mass. This equation of state assumes local thermodynamic equilibrium, which is valid in optically thick regions. Unless only a very small fraction of the debris remains near the binary, the debris is expected to very optically thick. As an order of magnitude estimate, the local optical depth to the midplane is roughly $\kappa M_{\star}/[2\uppi (2 a)^{2}]\simeq 10^{7} (M_{\star}/1\Msol)(a/0.01\AU)^{-2}$ where $\kappa = 0.34 \cm^{2}\gram^{-1}$ is the Thomson opacity. 

We adopt an artificial viscosity varying in the range between 0 and 1. The courant number is 0.3, as suggested for stable hydrodynamics evolution in \citep{Phantom}.

\subsection{Binary black holes}\label{subsec:binary}

We model the BBHs using two initially non-rotating sink particles. The sink particles interact only gravitationally with other sink particles and SPH particles. Furthermore, we assume that SPH particles that satisfy all of the following conditions accrete onto the sink particles:\\
\begin{itemize}
    \item the separation from the sink particle is less than $220r_{\rm g}$,\\
    \item the SPH particle is bound to the sink particle,\\
    \item the SPH particle is more bound to the sink particle than any other particle,\\
    \item the specific angular momentum of the SPH particle is smaller than that required to form a circular orbit at $r=220r_{\rm g}$.
\end{itemize}
However, if the separation from the sink particle is $\leq 180r_{\rm g}$, we assume that the SPH particle is accreted even though any of the above conditions is not satisfied. As expected, the accretion rate of a sink particle with these conditions is sensitive to the accretion radius. We found from additional test simulations for a few cases with different accretion radii between $200r_{\rm g}$  and $2500_{\rm g}$ that the accretion rate decreases with the accretion radius (roughly by a factor of ten between $200r_{\rm g}$  and $2500_{\rm g}$). The accretion radius is limited by the smoothing length. The smallest accretion radii allowed for the BH mass considered in our simulation are $\simeq 80 r_{\rm g}$ for $\Mbh=20\Msol$ and $\simeq 220 r_{\rm g}$  for $\Mbh=6\Msol$. For consistency, we choose the accretion radius to be $220r_{\rm g}$.  However, we also found from our test runs that the overall trend of the accretion rate and its modulation features are not significantly affected by resolution. Thus in our study we focus on some of those characteristic features of the accretion rate that are robust against the specific choice of the accretion radius. 

The momentum and energy associated with the particle accretion are properly taken into account. Because the addition or substraction of momentum due to close encounters and accretion could be significant in our simulations, we allow the BBHs to move freely and their orbit to evolve in response to the interactions with SPH particles. 

The binary black holes considered in our papers are hard binaries in the GW regime. The critical velocity of our binary-star system, typically defined for the binary-single scattering \footnote{The critical velocity $v_{\rm c}$ is defined as
\begin{equation}
v_{\rm c}^2 = \frac{Gm_1m_2(m_1+m_2+m_3)}{(m_1+m_2)m_3a},
\end{equation}
where $a$ is the semi-major axis of the binary, $m_1$ and $m_2$ are the mass of the objects in binary and $m_3$ is the mass of the intruder.} is $\simeq 2000$ km s$^{-1}$, which is  much larger than the relative velocity between the BBH and the star at infinity. Furthermore, the GW-driven merger timescales for our binaries are $\simeq 10^{3}-10^{10}\yr$, less than a Hubble time, which is depicted in Figure~\ref{fig:acc} and given in Table~\ref{tab:orbitchange}. We provide the initial parameters of the binaries in \S\ref{sub:initial} and Table~\ref{tab:initialparameter}.

\begin{figure}
\includegraphics[width=8.6cm]{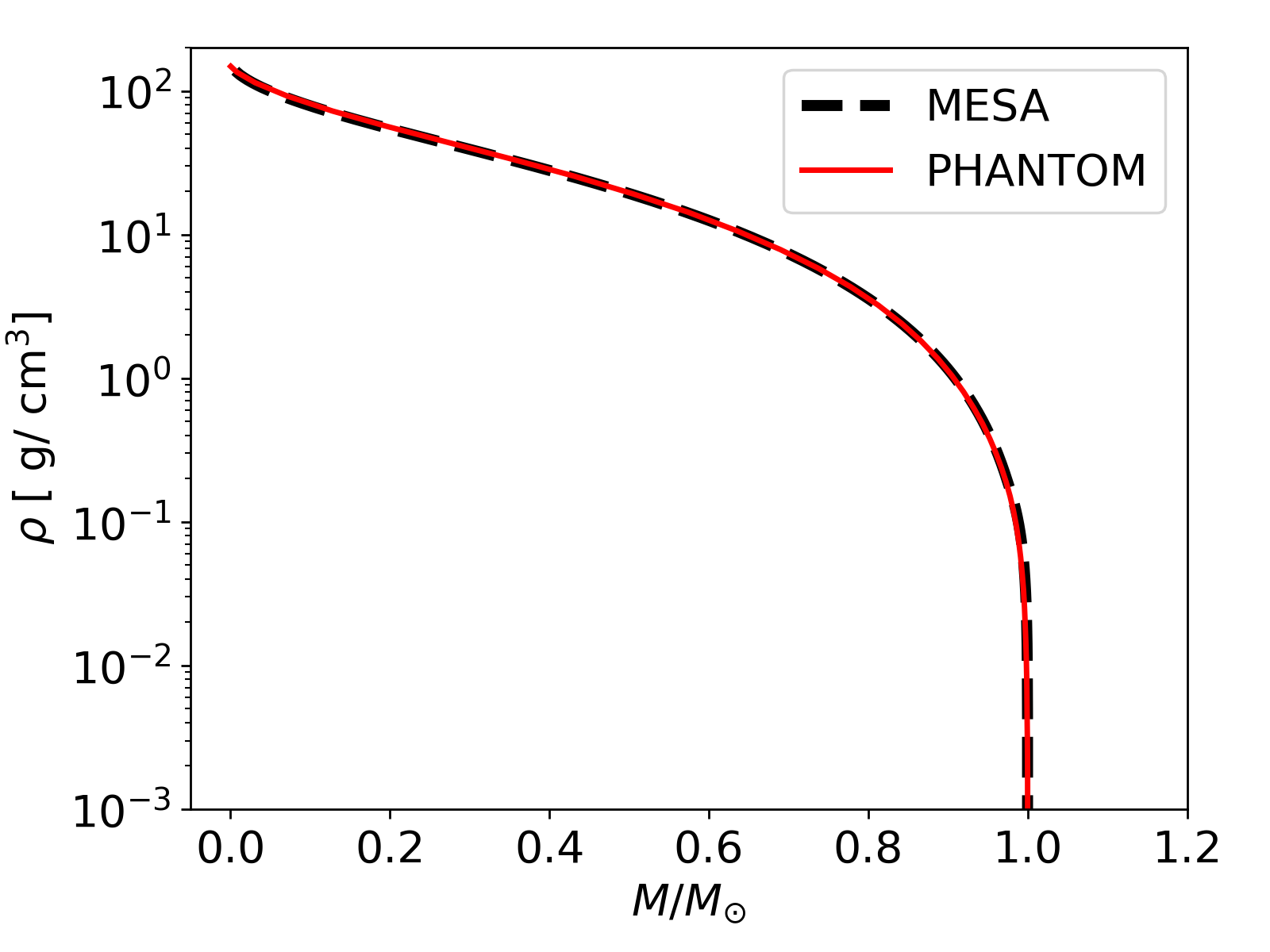}
	\caption{The density profiles of the MESA stellar model (dashed black) and of the star (red solid) mapped on our 3-D grid and relaxed for five stellar dynamical times, as a function of the fractional enclosed mass. }	\label{fig:relaxedstar} 
\end{figure}

\begin{figure*}
\includegraphics[width=8.3cm]{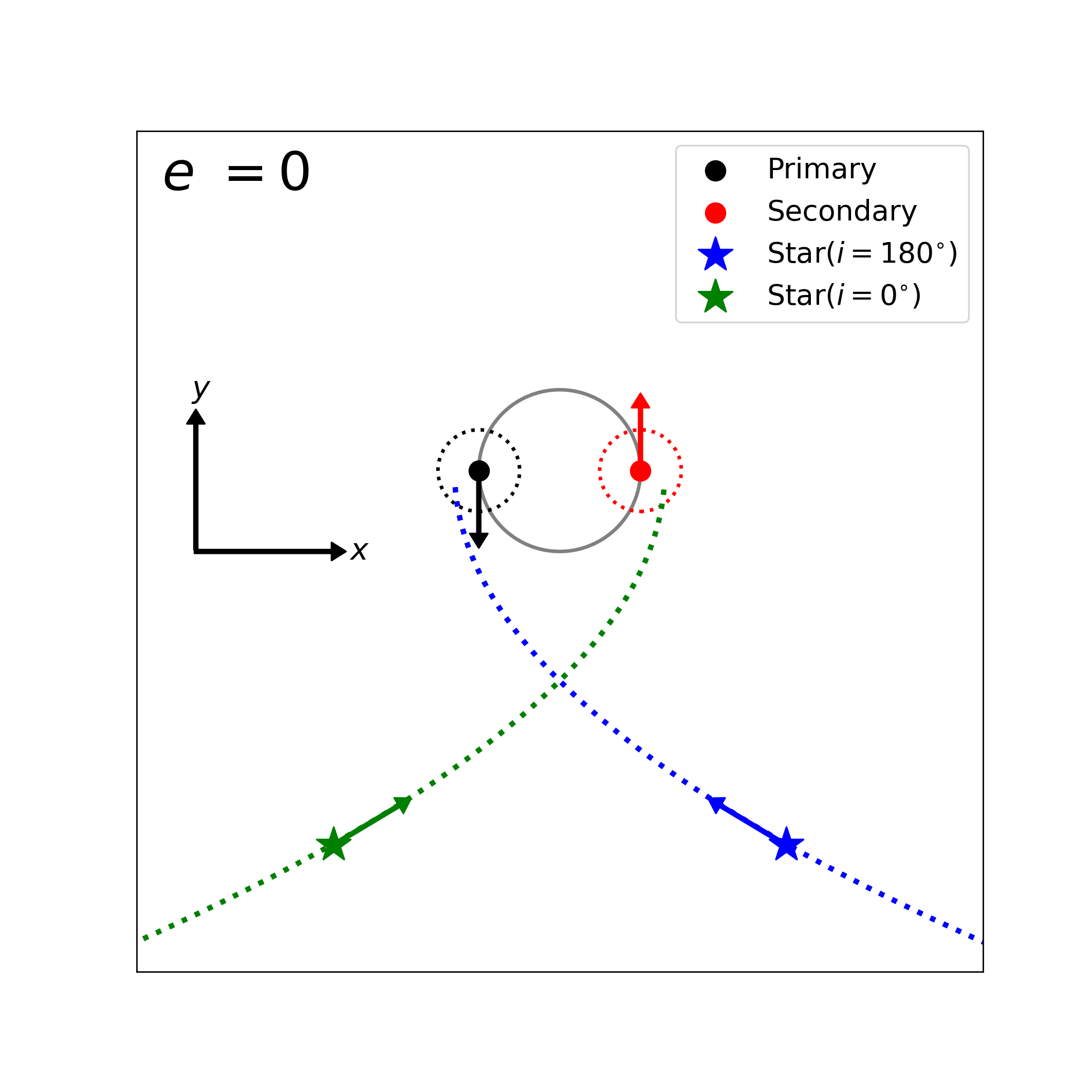}
\includegraphics[width=8.3cm]{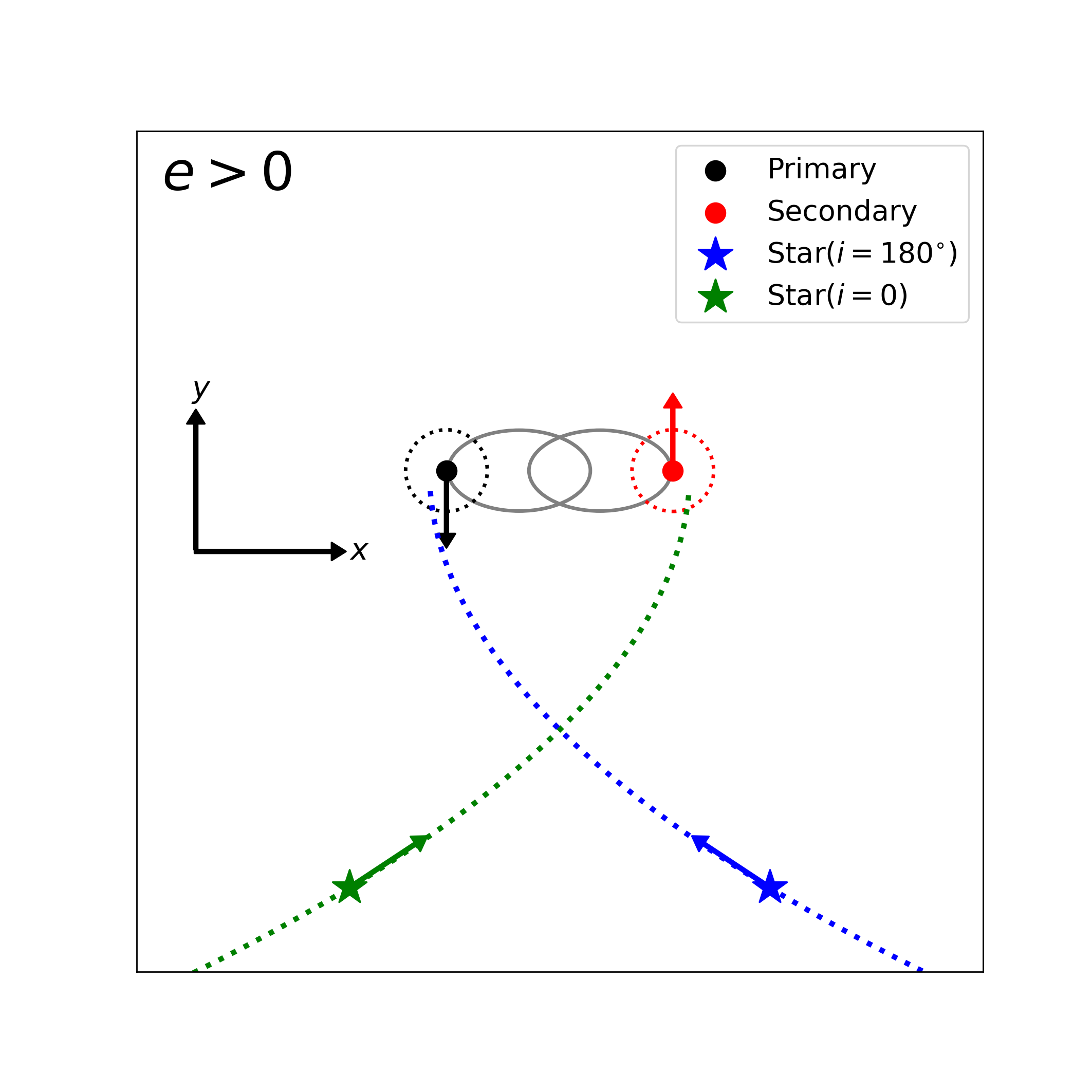}\\
\includegraphics[width=8.3cm]{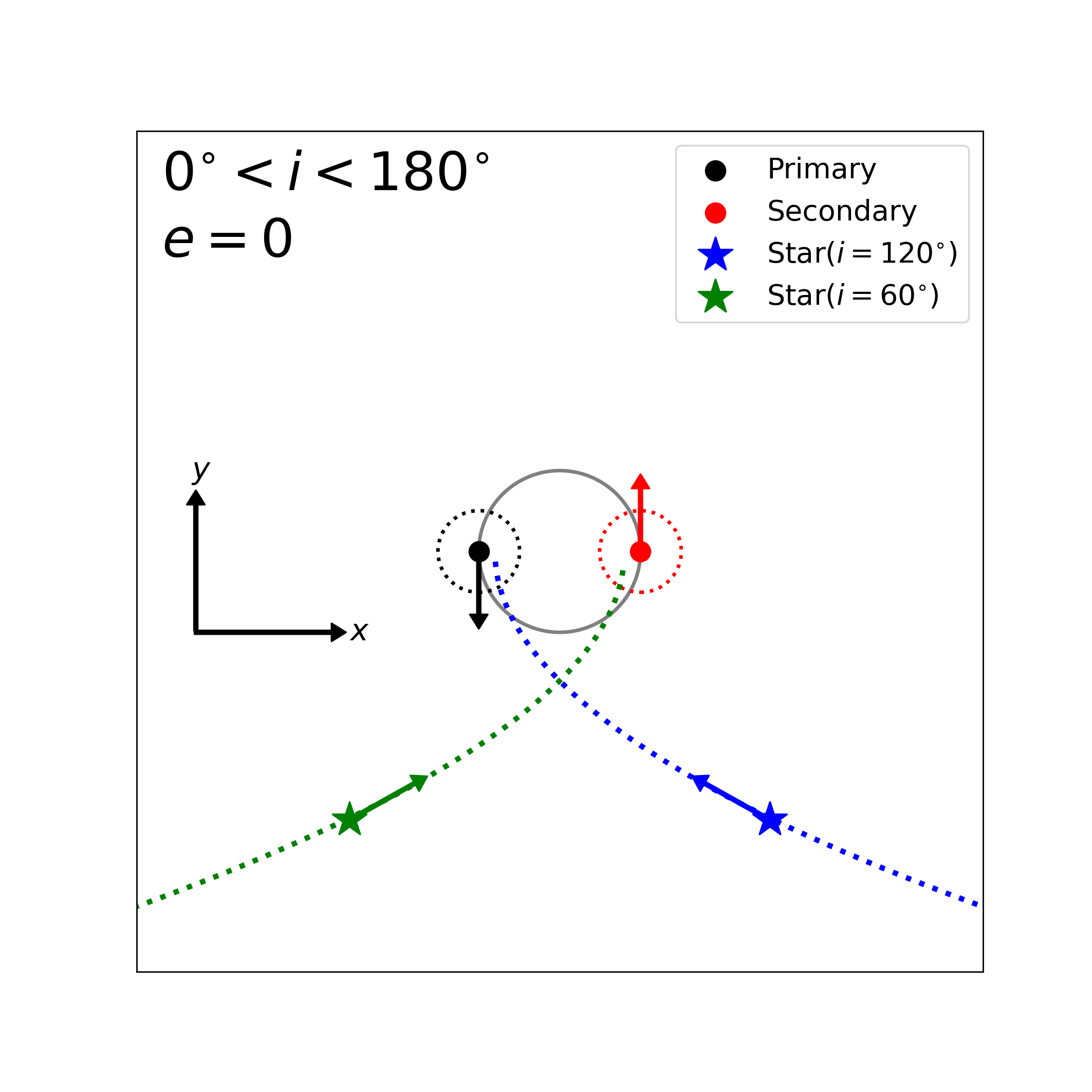}
\includegraphics[width=8.3cm]{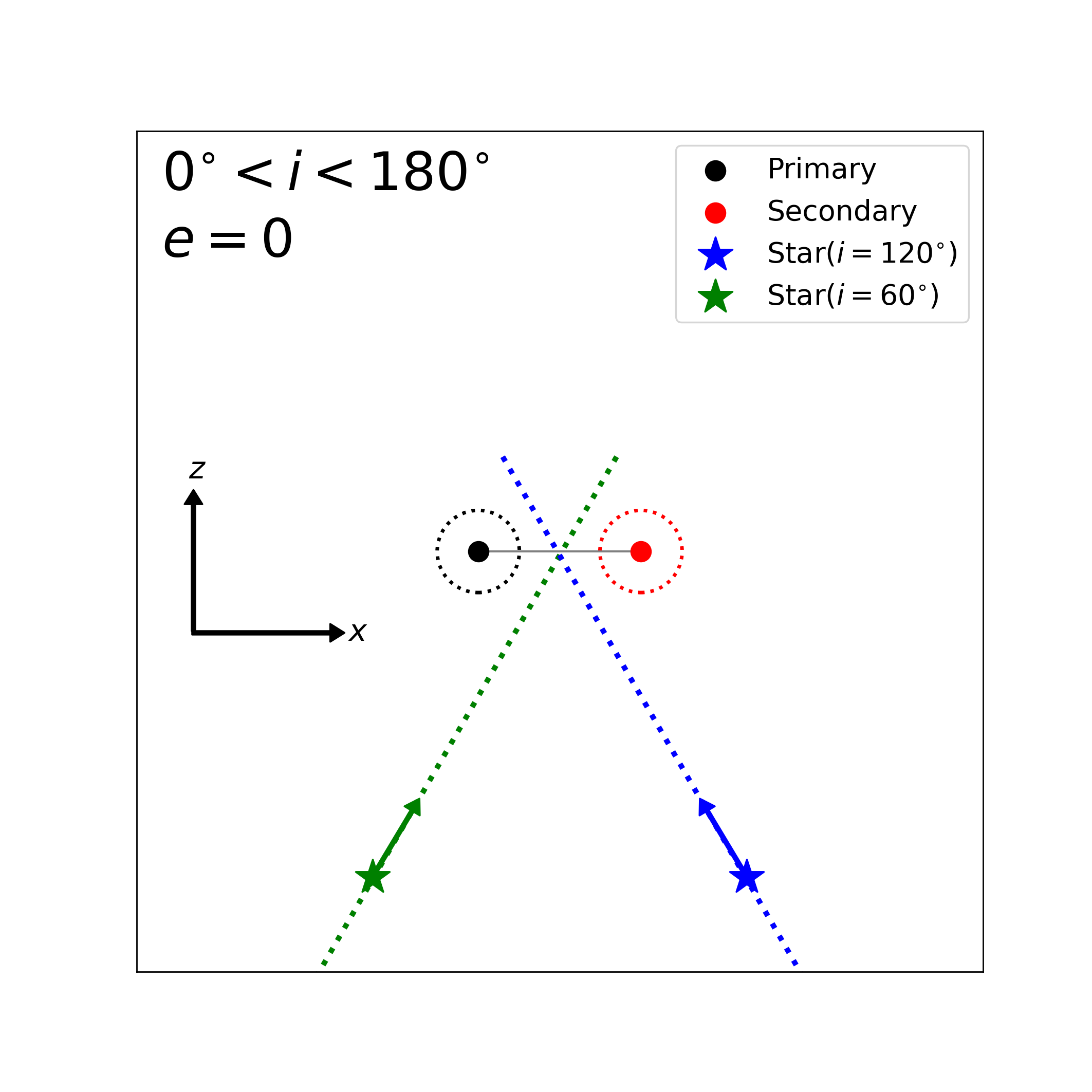}
	\caption{Schematic diagrams for the configuration of the binary (black circle : primary and red circle : secondary) and star (green star) at $t=0$ for three representative cases, 1) \textit{top-left} panel: $e=0$, 2) \textit{top-right} panel: $e>0$ and $i=0$ and 3) \textit{bottom} panels: $e=0$ and $0<i<180^{0}$. They are drawn not to scale. the \textit{left} and \textit{middle} panels show the orbits in the $x-y$ plane. The two \textit{bottom} panels illustrate the orbits in the $x-y$ (\textit{left}) and $x-z$ (\textit{right}) plane. The grey solid circle and ellipses depict the orbit of the binary, the green and blue dotted lines the trajectory of the star, and the red dotted circle the tidal radius of the secondary BH. The arrows indicate the instantaneous direction of motion. }	\label{fig:IC} 
\end{figure*}

\subsection{Stellar model}\label{subsec:star}
The initial state of the star is described using a stellar model for a $1\Msol$ middle-age main-sequence star evolved using the stellar evolution code {\small MESA} \citep{Paxton+2011}. We first map the $1-D$ {\small MESA} model onto our 3-D computational grid and relax it for a few stellar dynamical times ($\sqrt{R_{\star}^{3}/GM_{\star}}$) without including the gravity from the BBH. Figure~\ref{fig:relaxedstar} shows the density profile of the star relaxed for five stellar dynamical times (red solid) and the {\small MESA} model (black dashed) as a function of the fractional enclosed mass. We then use the fully relaxed star as our initial model in the simulations for close encounters. We note that, even in those cases, the initial distance of the relaxed star from the binary is sufficiently far ($\sim 5-10$ times the maximum of the tidal radius and the semimajor axis of the binary),  that the star has enough time to settle before it starts to be tidally deformed. 

The star is modelled with $N_{\rm SPH}=5\times 10^{5}$ equal-mass SPH particles. We performed additional test simulations for the equal-mass binary with the accretion radius of $100r_{\rm g}$ using three different resolutions, $N_{\rm SPH}=10^{5}, 5\times10^{5}$ and $10^{6}$. We find that our results for the orbit evolution of the binary are in excellent agreement among those simulations with different resolutions. As we will show in \S\ref{subsec:accretion}, the accretion rate is the highest at the first peak and then decreases over time. The accretion rate at the first peak has not been converged within this resolution range: it is roughly a factor of 2 smaller for a resolution lower by a factor of 5. However, we found that the accretion rate typically converges as the rate decreases. 

Hereafter, the orbital parameters with (without) the subscript $\star$ indicate those for the stellar orbit relative to the binary (binary orbit). 

%The polytropic approximation for the $1\Msol$ star has been widely used in a number of previous works for TDEs \citep[e.g.,][]{LuminetCarter1986,Guillochon2013,Mainetti+2017}. The recent studies adopting realistic stellar internal structure in simulations for TDEs \citep[e.g.,][]{Ryu+2020a} show that there are mainly two important differences in the TDE outcomes in their simulations, compared to those with a $1\Msol$ polytropic star with $\gamma=5/3$ and those from a star with realistic internal structure. 

\subsection{Initial conditions and parameters}\label{sub:initial}

The black holes are initially located along the $x-$axis; the orbital rotation axis is parallel to $+\textbf{z}$ in all of our models, as illustrated in Figure~\ref{fig:IC}. The star approaches the binary on a parabolic orbit ($1-e_{\star}=10^{-6}$), from a distance from which the travel time to the pericenter (assuming the binary is a point particle with mass $\Mbhp+\Mbhs$) is $t_{\rm arr}= \upsilon P$. Here, $\upsilon$ is a constant factor and $P$ is the initial orbital period of the binary. We take $\upsilon = 2$ in all our simulations with $a=0.01\AU$ except for the models with different phase angles at the first encounter (model names starting with $\Delta\upsilon$). We use $\upsilon=1$ for the simulations with $a=0.03$ and $0.1\AU$. The initial separation is more than five times greater than the larger value of the binary semimajor axis $a$ and the nominal tidal radius of the binary. 

The pericenter distance of the initial stellar orbit relative to the binary is $r_{\rm p,\star}=0.5(1 + e)a + r_{\rm t}/\alpha$. Again, $r_{\rm t}$ is the nominal tidal radius of the disruptor BH and $\alpha$ is a constant factor which determines how close the star approaches the disruptor BH. Given that the travel time is an integer multiple of $P$, if $r_{\rm p,\star}$ is sufficiently large (say $r_{\rm p,\star}> 2a$), the BHs and the star would be aligned along the $x-$axis by the time the star arrives at the pericenter. For this case, the separation between the disruptor BH and star would be $\sim r_{\rm t,\star}/\alpha$ and $\alpha$ is equivalent to the penetration factor commonly used in TDE studies. However, in our simulations where $r_{\rm p,\star}\lesssim 1.7a$, the star's orbit within $r\lesssim 2a$ would become significantly deviated from the original orbit because of the potential of the disruptor BH, not the binary's potential, starts to govern the motion of the star.

These initial conditions allow us to simulate astrophysically probable scenarios in a consistent way. Alternatively, we could have chosen initial conditions that yield a uniform configuration at the first closest encounter in order to decouple the impact of one parameter from the other. However, because the encounter parameters (e.g., the angle between the velocity vectors of the star and disruptor BH and their relative speed) are closely correlated, changing one parameter often leads to a modification of other parameters. In most cases, such an idealized setup leads to situations which are not necessarily astrophysically plausible. 

With our set of initial conditions, we investigate the impact of close encounters on the evolution of BBHs with different initial binary orbital parameters, 1) $a= 0.01$, $0.03$ and $0.1\AU$, 2) $q=1$ and $0.3$, 3) $e=0$, $0.5$ and $0.9$. For the unequal-mass case, we consider two scenarios, one in which the primary (i.e., more massive) BH is the disruptor and one in which the secondary (i.e., less massive) BH is the disruptor. In each simulation, we change only one of those parameters from those considered in our fiducial model, which is an in-plane prograde ($i=0$) encounter between a circular equal-mass BBH with  $\Mbhp=\Mbhs=20\Msol$ and $a=0.01\AU$, and the star. In addition, we perform the same set of simulations for in-plane retrograde encounters ($i=180^{\circ}$). The in-plane encounters can occur in circumstances where binaries and stars orbit in a common plane such as AGN disks. However, in clusters, the inclination angle between the two orbital planes can be large. Motivated by this, we perform other simulations with inclinations of $i=60^{\circ}$ and $120^{\circ}$, while the rest of the parameters are the same as for the fiducial model. We change the inclination only by rotating the orbital plane of the star around the $y$-axis. The initial configuration for the cases with $i=60^{\circ}$ and $120^{\circ}$ are depicted in the bottom panels of Figure~\ref{fig:IC}. 

Lastly, we also study the dependence on $\alpha$ and $t_{\rm arr}$. In particular, a different arrival time means that the first closest encounter occurs at a different location in the orbit. This is motivated by the fact that the star and binary's orbits are not necessarily synchronized, but rather the phase of the two orbits is arbitrary. To simulate the encounters with different phase angles, we initially place the star gradually farther away from the binary so that the star arrives at a different time. We consider five different arrival times, $\Delta\upsilon=v-2= 1/12$, $2/12$, $3/12$, $4/12$ and $5/12$ while the rest of the parameters remain the same as the fiducial case. 

Each of the models is integrated up to $\simeq 15 P$. Note that the orbital parameters of the binary after the first encounter typically settle at $t/P\lesssim 5$ and remain nearly constant afterwards.

\begin{figure*}
\includegraphics[width=8.6cm]{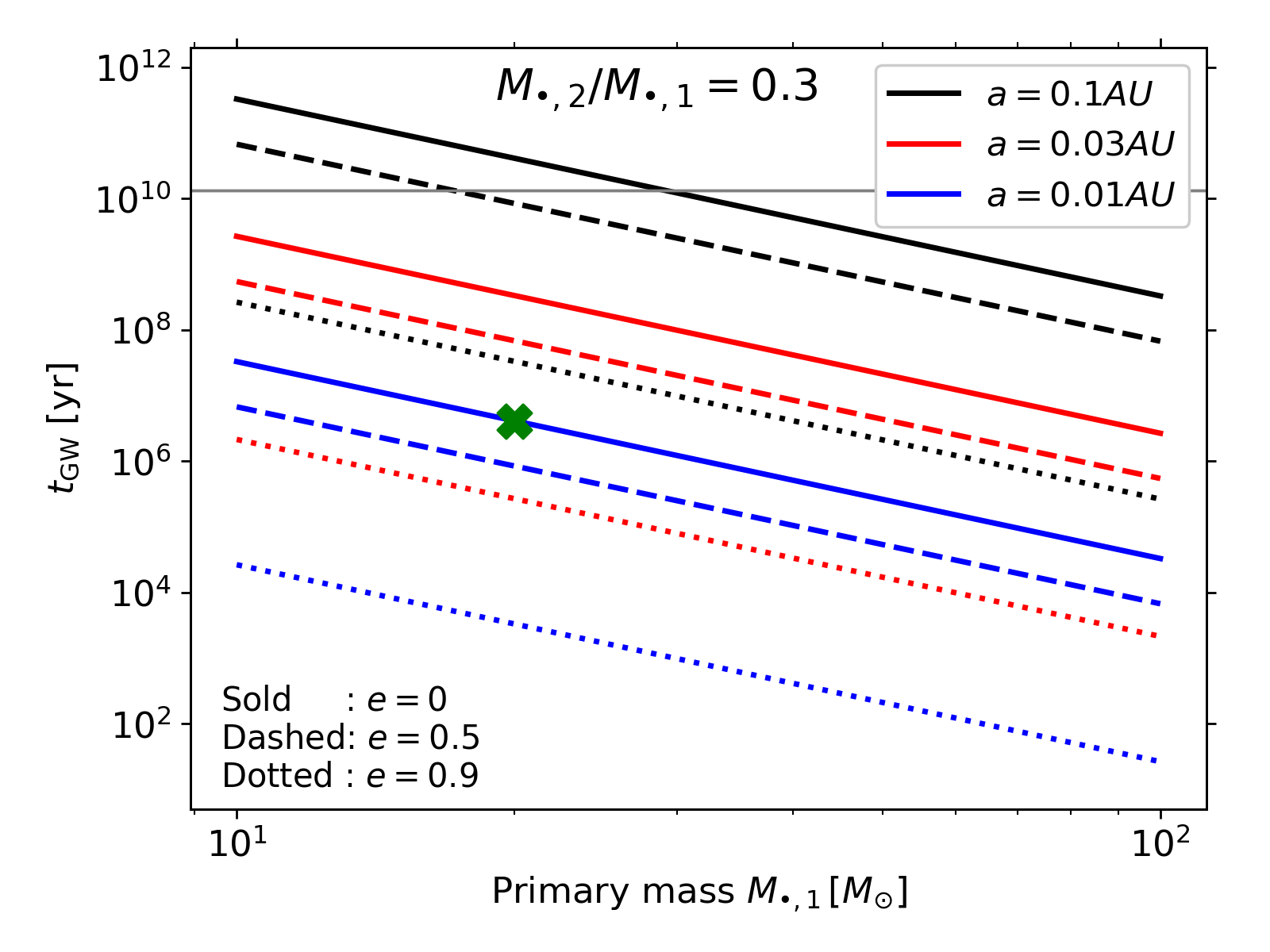}
\includegraphics[width=8.6cm]{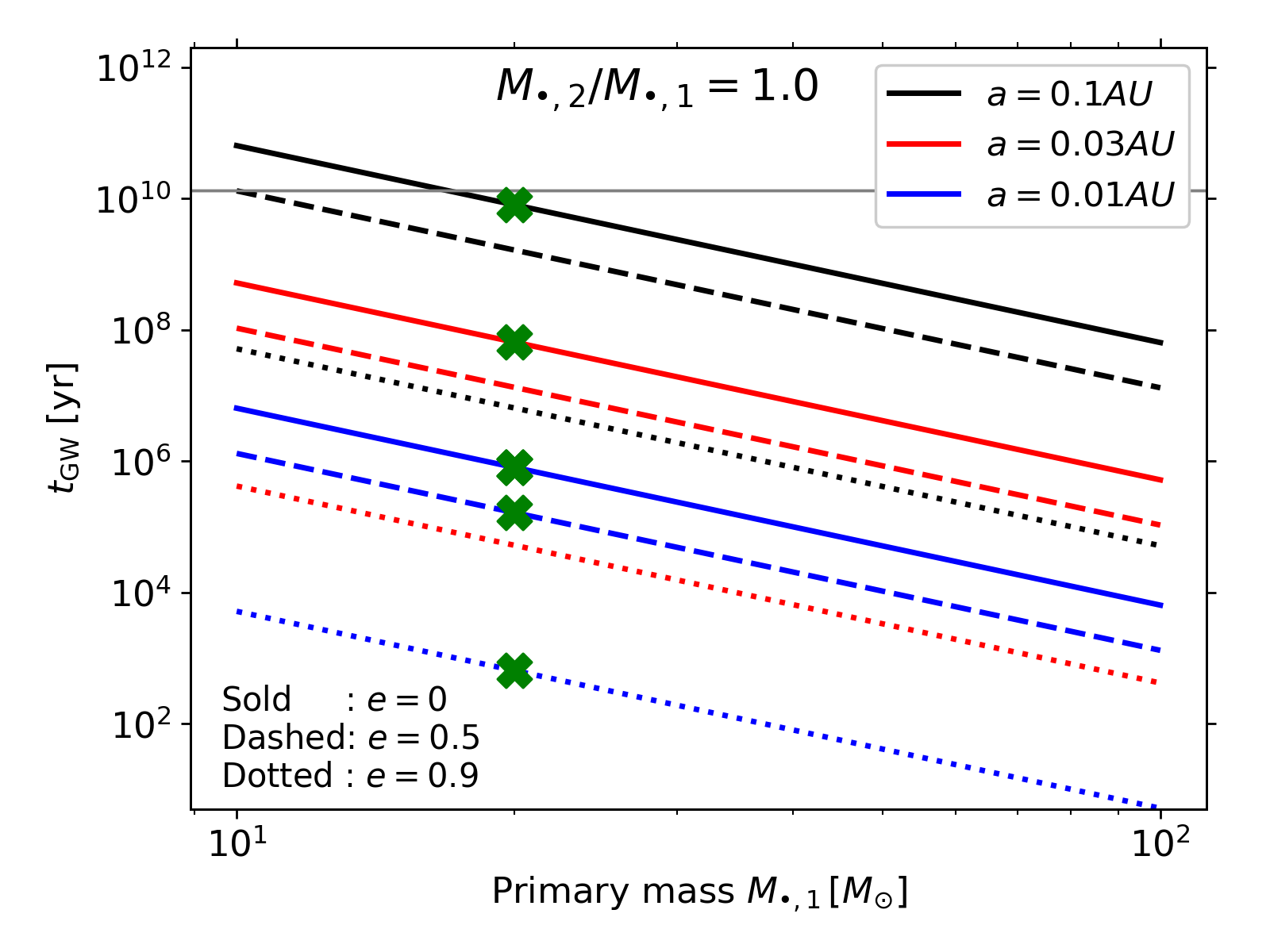}
	\caption{The GW-driven merger timescale as a function of the primary BH mass $\Mbhp$ in binaries with $a=0.1$ (black), $0.03$ (red) and $0.01$ (blue) for $q=0.3$ (\textit{left}) and $q$ (\textit{right}). Different eccentricities are indicated using a different linestyle: $e=0$ (solid), $0.5$ (dashed) and $0.9$ (Dotted). We mark $t_{\rm GW}$ for the initial orbital parameters of our model binaries using green crosses.  }	\label{fig:GWt} 
\end{figure*}

To provide an insight of which regime the BBHs considered in our simulations fall, we show the gravitational wave(GW)-driven merger timescale $t_{\rm GW}$ as a function of $\Mbhp$ in Figure~\ref{fig:GWt}. We mark the time scale for the initial binary parameters using green crosses. Most of the binaries considered are compact enough for the GW emission to drive them to merge in less than $10^{6}\yr$. On the other hand, the widest binary with $a=0.1\AU$ has a merger time scale comparable to a Hubble time. 

%Phase

We summarize the initial parameters considered in our simulations in Table~\ref{tab:initialparameter}.

\begin{table}
\begin{tabular}{  c c c c c c c c} 
\hline
Model name & $M_{\rm d}$ & $q$ & $a[{\rm AU}]$ & $e$& $i~[^{\circ}]$  & $\Delta\upsilon$  & $\alpha$ \\
\hline
$\alpha1-pro.$ & 20 & 1 & 0.01 & 0.0 & 0 & 0 & 1 \\
$\alpha2-pro.$ & 20 & 1 & 0.01 & 0.0 & 0 & 0 & 2 \\
$\alpha3-pro.$ & 20 & 1 & 0.01 & 0.0 & 0 & 0 & 3 \\
$Fiducial$ & 20 & 1 & 0.01 & 0.0 & 0 & 0 & 4 \\
$\alpha5-pro.$ & 20 & 1 & 0.01 & 0.0 & 0 & 0 & 5 \\
$\alpha1-retro.$ & 20 & 1 & 0.01 & 0.0 &  180 &0 & 1 \\
$\alpha2-retro.$ & 20 & 1 & 0.01 & 0.0 &180 &  0 & 2 \\
$\alpha3-retro.$ & 20 & 1 & 0.01 & 0.0 & 180 & 0 & 3 \\
$\alpha4-retro.$ & 20 & 1 & 0.01 & 0.0 & 180 & 0 & 4 \\
$\alpha5-retro.$ & 20 & 1 & 0.01 & 0.0 &180 &  0 & 5 \\
\hline
$a0.03-pro.$ & 20 & 1 & 0.03 & 0.0 & 0 & 0 & 4  \\
$a0.03-retro.$ & 20 & 1 & 0.03 & 0.0 & 180 & 0 & 4  \\
$a0.1-pro.$ & 20 & 1 & 0.1 & 0.0 & 0 & 0 & 4  \\
$a0.1-retro.$ & 20 & 1 & 0.1 & 0.0 & 180 & 0 & 4  \\
\hline
$q0.3P-pro.$ & 20 & 0.3 & 0.01 & 0.0 & 0 & 0 & 4  \\
$q0.3S-pro.$ & 6 & 0.3 & 0.01 & 0.0 & 0 & 0 & 4  \\
$q0.3P-retro.$ & 20 & 0.3 & 0.01 & 0.0 & 180 & 0 & 4  \\
$q0.3S-retro.$ & 6 & 0.3 & 0.01 & 0.0 & 180 & 0 & 4  \\
\hline
$e0.5-pro.$ & 20 & 1 & 0.01 & 0.5 & 0 & 0 & 4  \\
$e0.5-retro.$ & 20 & 1 & 0.01 & 0.5 & 180 & 0 & 4  \\
$e0.9-pro.$ & 20 & 1 & 0.01 & 0.9 & 0 & 0 & 4  \\
$e0.9-retro.$ & 20 & 1 & 0.01 & 0.9 & 180 & 0 & 4  \\
\hline
$i60$ & 20 & 1 & 0.01 & 0.0 & 60 & 0 & 4 \\
$i120$ & 20 & 1 & 0.01 & 0.0 & 120 & 0 & 4 \\
\hline
$\Delta\upsilon1/12$ & 20 & 1 & 0.01 & 0.0 & 0 & 1/12& 4  \\
$\Delta\upsilon2/12$ & 20 & 1 & 0.01 & 0.0 & 0 &2/12 &  4 \\
$\Delta\upsilon3/12$ & 20 & 1 & 0.01 & 0.0 & 0  &3/12 &  4 \\
$\Delta\upsilon4/12$ & 20 & 1 & 0.01 & 0.0 & 0 &  4/12 &4  \\
$\Delta\upsilon5/12$ & 20 & 1 & 0.01 & 0.0 & 0 &5/12 &  4  \\
\hline
\end{tabular}
\caption{The initial model parameters. $M_{\rm d}$ is the mass of the BH that disrupts the star (disruptor BH), $q=\Mbhs/\Mbhp$,  $a$ the semimajor axis of the binary, $e$ the eccentricity of the binary at the first encounter, $i$ the inclination angle between the binary orbital plane and stellar orbital plane and $\Delta\upsilon$ the difference in the arrival time from that for the fiducial model ($\upsilon=2$). The pericenter distance of the incoming orbit is $r_{\rm p,\star}=0.5(1 + e)a + r_{\rm t}/\alpha$ where $r_{\rm t}$ is the nominal tidal radius of the disruptor BH. In all cases, $M_{\rm d}= \Mbhp$ except in Models 7 and 12 where $M_{\rm d}= \Mbhs$. The orbital period of the equal-mass binary is $\sim44$ hours for $a=0.1\AU$, $\sim 7$ hours for $a=0.03\AU$ and $\sim1.4$ hours for $a=0.01\AU$}\label{tab:initialparameter}
\end{table}

\begin{table*}
\begin{tabular}{  c c c c c c c c c c c c} 
\hline
Model name & $a_{0}[\AU]$ & $a[\AU]$ & $\Delta a / a_{0}$ & $e_0$ & $e$ & $\Delta e$ & $t_{\rm GW,0}[\yr]$   & $t_{\rm GW}[\yr]$  & $\Delta t_{\rm GW}/t_{\rm GW,0}$ & Outcome & Type\\
\hline
$\alpha1-pro.$ &  0.01 & 9.94e-03 & -0.006 &   0.0 & 0.058 & 0.027 &  8.03e+05 & 7.66e+05 & -0.047   & PTDE & Regular \\
$\alpha2-pro.$ &  0.01 & 1.20e-02 & 0.202 &   0.0 & 0.117 & 0.045 &  8.03e+05 & 1.52e+06 & 0.894   & FTDE & Non-regular \\
$\alpha3-pro.$ &  0.01 & 1.14e-02 & 0.136 &   0.0 & 0.138 & 0.056 &  8.03e+05 & 1.17e+06 & 0.458   & FTDE & Non-regular \\
$Fiducial$ &  0.01 & 1.12e-02 & 0.121 &   0.0 & 0.134 & 0.055 &  8.03e+05 & 1.12e+06 & 0.393   & FTDE & Non-regular \\
$\alpha5-pro.$ &  0.01 & 1.11e-02 & 0.109 &   0.0 & 0.122 & 0.051 &  8.03e+05 & 1.09e+06 & 0.361   & FTDE & Non-regular \\
$\alpha1-retro.$ &  0.01 & 1.00e-02 & -0.000 &   0.0 & 0.003 & 0.001 &  8.03e+05 & 8.02e+05 & -0.001   & PTDE & Regular \\
$\alpha2-retro.$ &  0.01 & 9.96e-03 & -0.004 &   0.0 & 0.016 & 0.008 &  8.03e+05 & 7.88e+05 & -0.019   & PTDE & Regular \\
$\alpha3-retro.$ &  0.01 & 9.90e-03 & -0.010 &   0.0 & 0.032 & 0.015 &  8.03e+05 & 7.65e+05 & -0.048   & FTDE & Regular \\
$\alpha4-retro.$ &  0.01 & 9.77e-03 & -0.023 &   0.0 & 0.049 & 0.024 &  8.03e+05 & 7.20e+05 & -0.103   & FTDE & Regular \\
$\alpha5-retro.$ &  0.01 & 9.61e-03 & -0.040 &   0.0 & 0.066 & 0.032 &  8.03e+05 & 6.63e+05 & -0.174   & FTDE & Regular \\
$a0.03-pro.$ &  0.03 & 3.15e-02 & 0.049 &   0.0 & 0.080 & 0.012 &  6.51e+07 & 7.56e+07 & 0.161   & FTDE & Regular \\
$a0.03-retro.$ &  0.03 & 2.56e-02 & -0.146 &   0.0 & 0.212 & 0.039 &  6.51e+07 & 2.58e+07 & -0.603   & FTDE & Regular \\
$a0.1-pro.$ &  0.10 & 1.08e-01 & 0.076 &   0.0 & 0.022 & 0.001 &  8.03e+09 & 1.07e+10 & 0.336   & PTDE & Regular \\
$a0.1-retro.$ &  0.10 & 7.91e-02 & -0.209 &   0.0 & 0.301 & 0.018 &  8.03e+09 & 1.77e+09 & -0.780   & PTDE & Regular \\
$q0.3P-pro.$ &  0.01 & 1.14e-02 & 0.142 &   0.0 & 0.080 & 0.032 &  4.12e+06 & 6.69e+06 & 0.625   & FTDE & Regular \\
$q0.3S-pro.$ &  0.01 & 1.12e-02 & 0.123 &   0.0 & 0.097 & 0.040 &  4.12e+06 & 5.90e+06 & 0.432   & FTDE & Non-regular \\
$q0.3P-retro.$ &  0.01 & 9.42e-03 & -0.058 &   0.0 & 0.073 & 0.036 &  4.12e+06 & 3.12e+06 & -0.242   & FTDE & Regular \\
$q0.3S-retro.$ &  0.01 & 8.10e-03 & -0.190 &   0.0 & 0.232 & 0.133 &  4.12e+06 & 1.24e+06 & -0.700   & FTDE & Non-regular \\
$e0.5-pro.$ &  0.01 & 1.07e-02 & 0.066 &   0.5 & 0.446 & -0.024 &  1.64e+05 & 2.94e+05 & 0.785   & FTDE & Non-regular \\
$e0.5-retro.$ &  0.01 & 9.83e-03 & -0.017 &   0.5 & 0.518 & 0.009 &  1.64e+05 & 1.36e+05 & -0.174   & FTDE & Regular \\
$e0.9-pro.$ &  0.01 & 1.02e-02 & 0.022 &   0.9 & 0.875 & -0.011 &  6.46e+02 & 1.52e+03 & 1.350   & FTDE & Non-regular \\
$e0.9-retro.$ &  0.01 & 9.93e-03 & -0.007 &   0.9 & 0.913 & 0.006 &  6.46e+02 & 3.91e+02 & -0.395   & FTDE & Regular \\
$i60$ &  0.01 & 1.01e-02 & 0.006 &   0.0 & 0.068 & 0.031 &  8.03e+05 & 7.98e+05 & -0.007   & FTDE & Regular \\
$i120$ &  0.01 & 9.88e-03 & -0.013 &   0.0 & 0.047 & 0.022 &  8.03e+05 & 7.52e+05 & -0.064   & PTDE & Regular \\
$\Delta\upsilon1/12$ &  0.01 & 9.20e-03 & -0.080 &   0.0 & 0.068 & 0.034 &  8.03e+05 & 5.57e+05 & -0.307   & PTDE & Regular \\
$\Delta\upsilon2/12$ &  0.01 & 9.37e-03 & -0.063 &   0.0 & 0.053 & 0.026 &  8.03e+05 & 6.08e+05 & -0.243   & PTDE & Regular \\
$\Delta\upsilon3/12$ &  0.01 & 1.17e-02 & 0.168 &   0.0 & 0.102 & 0.041 &  8.03e+05 & 1.38e+06 & 0.721   & PTDE & Regular \\
$\Delta\upsilon4/12$ &  0.01 & 1.06e-02 & 0.057 &   0.0 & 0.104 & 0.046 &  8.03e+05 & 9.33e+05 & 0.161   & FTDE & Non-regular \\
$\Delta\upsilon5/12$ &  0.01 & 1.17e-02 & 0.168 &   0.0 & 0.147 & 0.059 &  8.03e+05 & 1.27e+06 & 0.578   & FTDE & Non-regular \\\hline
\end{tabular}
\caption{The changes in the orbital parameters, $a$ and $e$, due to close encounters. $a$, $e$ and $t_{\rm GW}$ are respectively the semimajor axis, eccentricity and GW-driven merger times of the binary after encounter. Those with the subscript '$_{0}$' indicate the values for the initial binary. $\Delta$ represents the change of a given quantity relative to its initial value (e.g., $\Delta a = a  - a_{0}$). The last two columns indicate the outcome of the encounter after the first encounter (PTDE: partial destruction and  FTDE: full destruction) and the type of stream evolution (see the text in \S\ref{subsec:overview}).}
\label{tab:orbitchange}
\end{table*}

\begin{figure*}

\includegraphics[width=4.3cm]{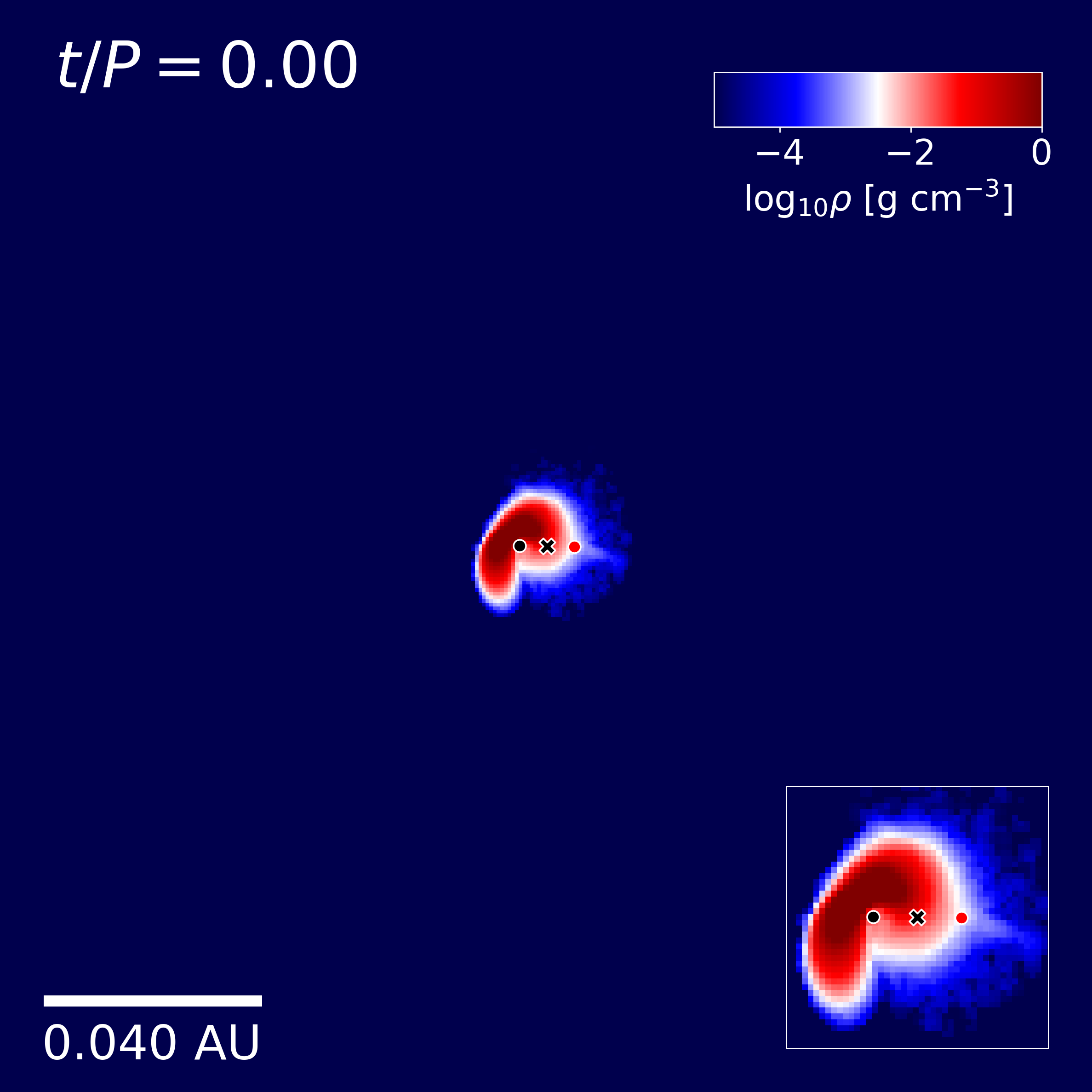}
\includegraphics[width=4.3cm]{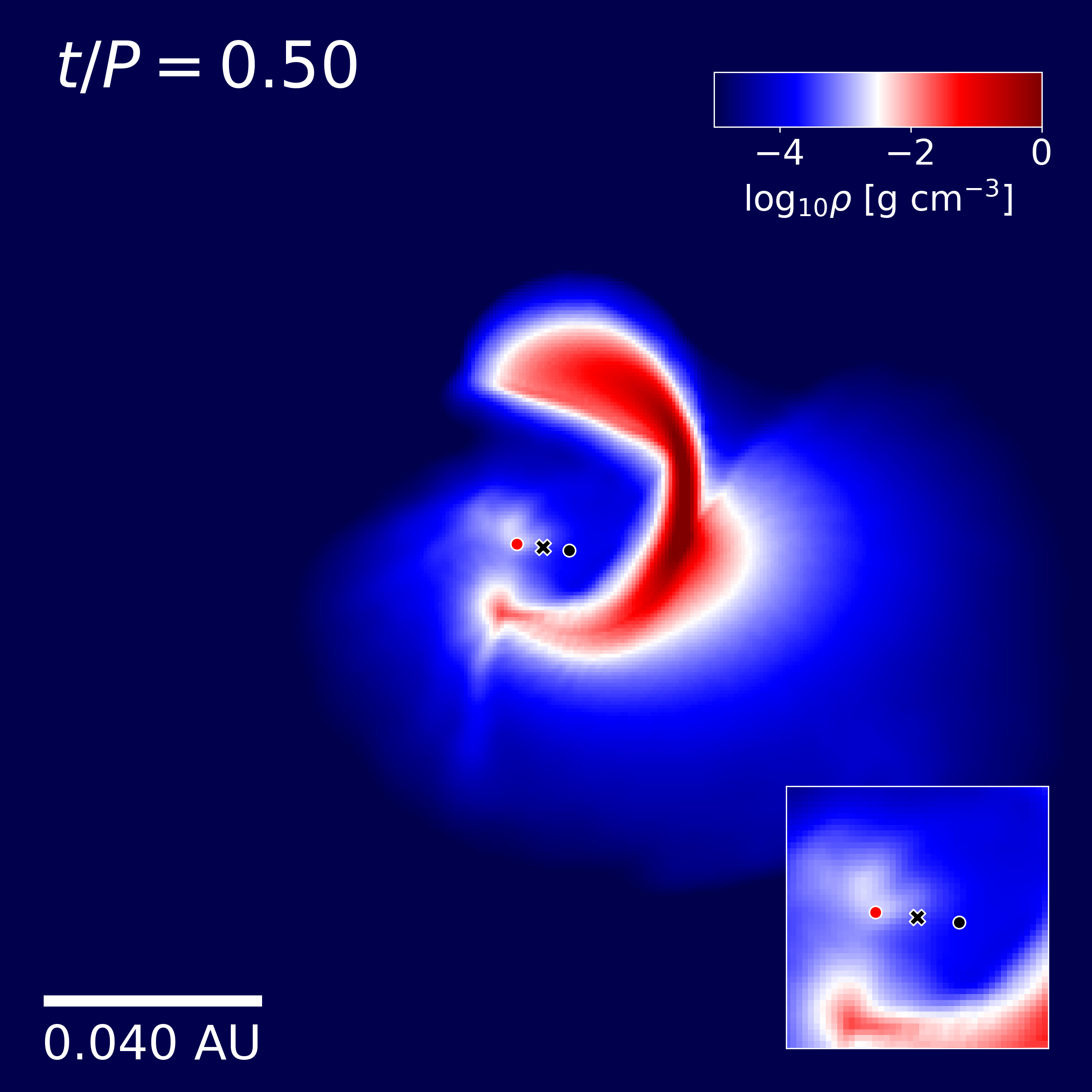}
\includegraphics[width=4.3cm]{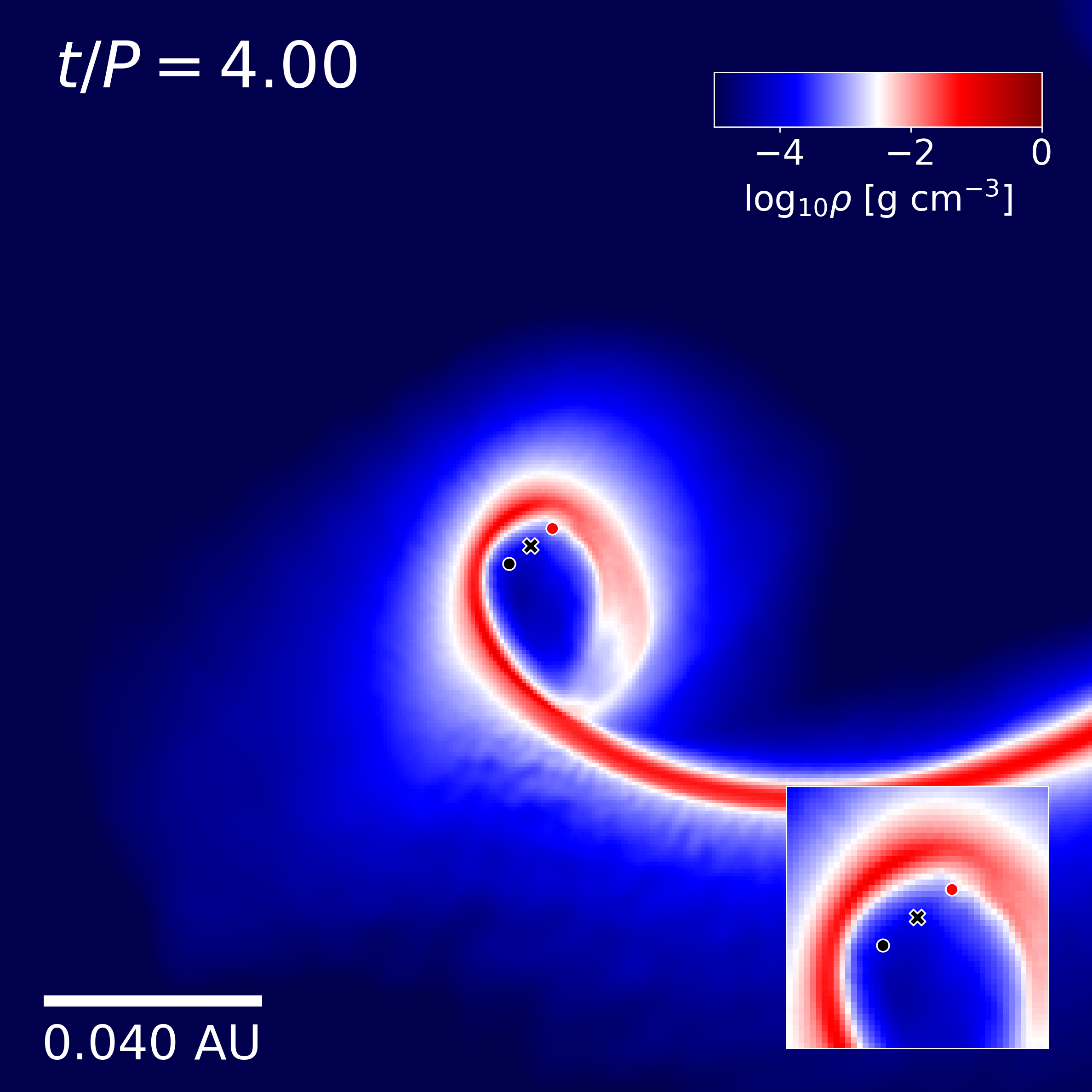}
\includegraphics[width=4.3cm]{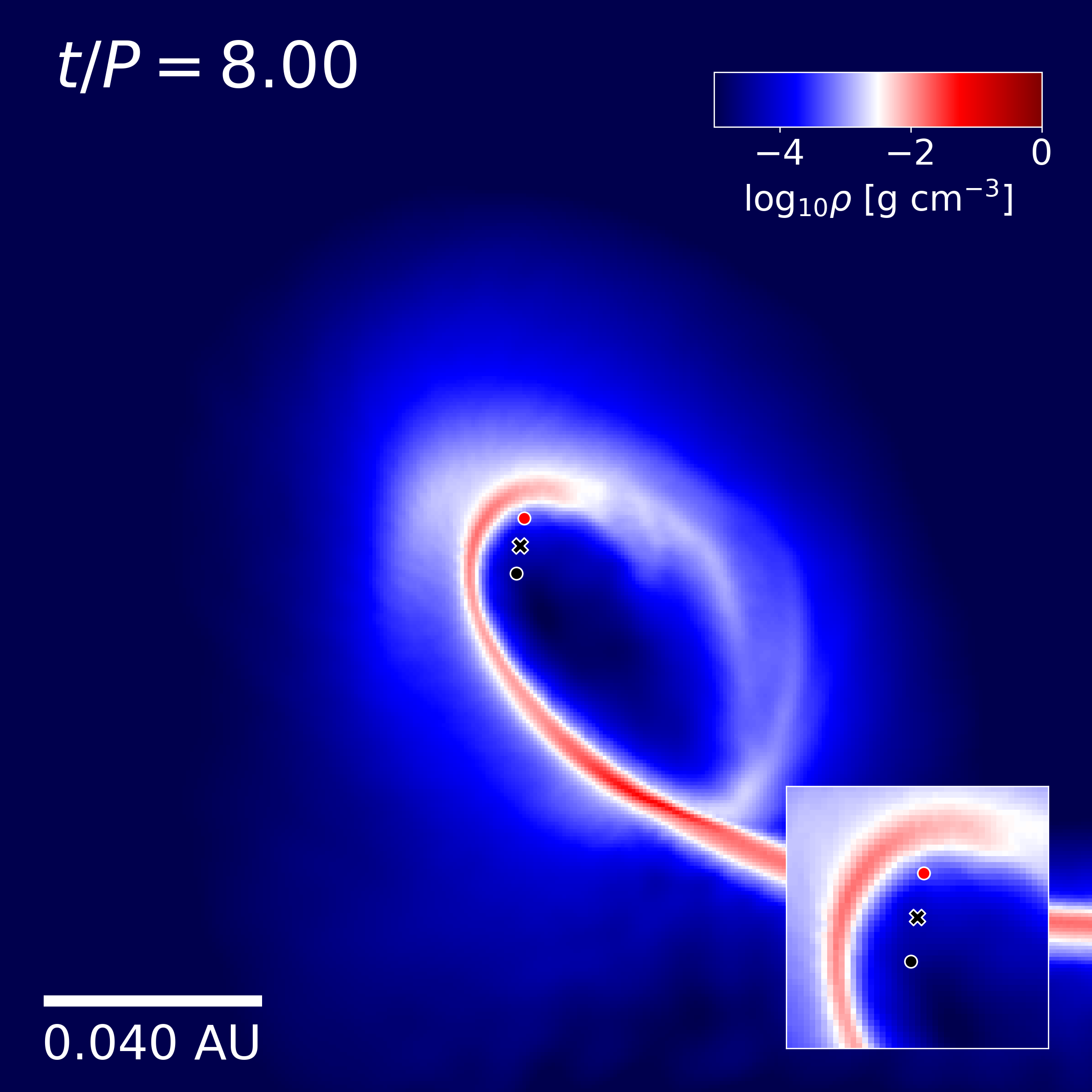}\\
\includegraphics[width=4.3cm]{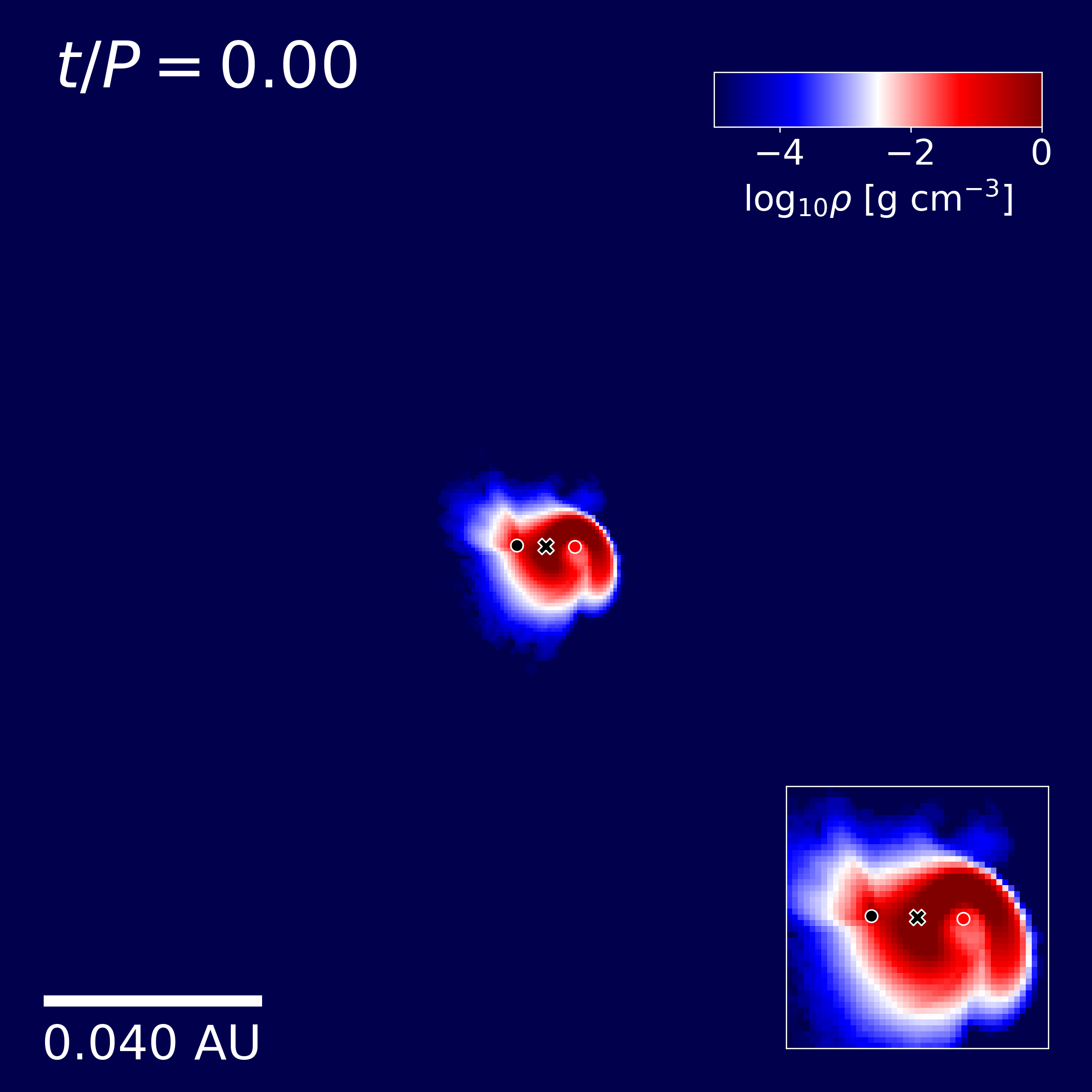}
\includegraphics[width=4.3cm]{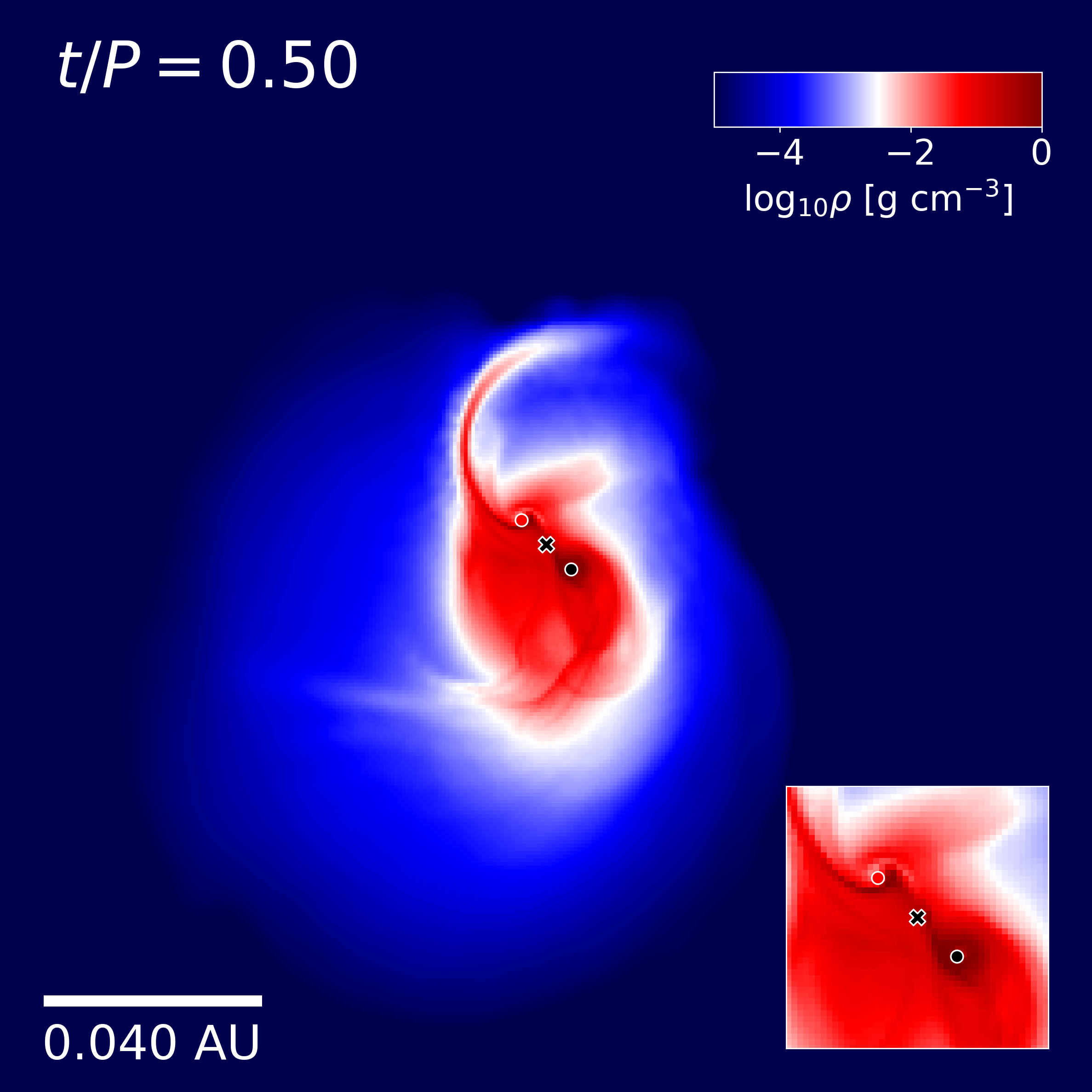}
\includegraphics[width=4.3cm]{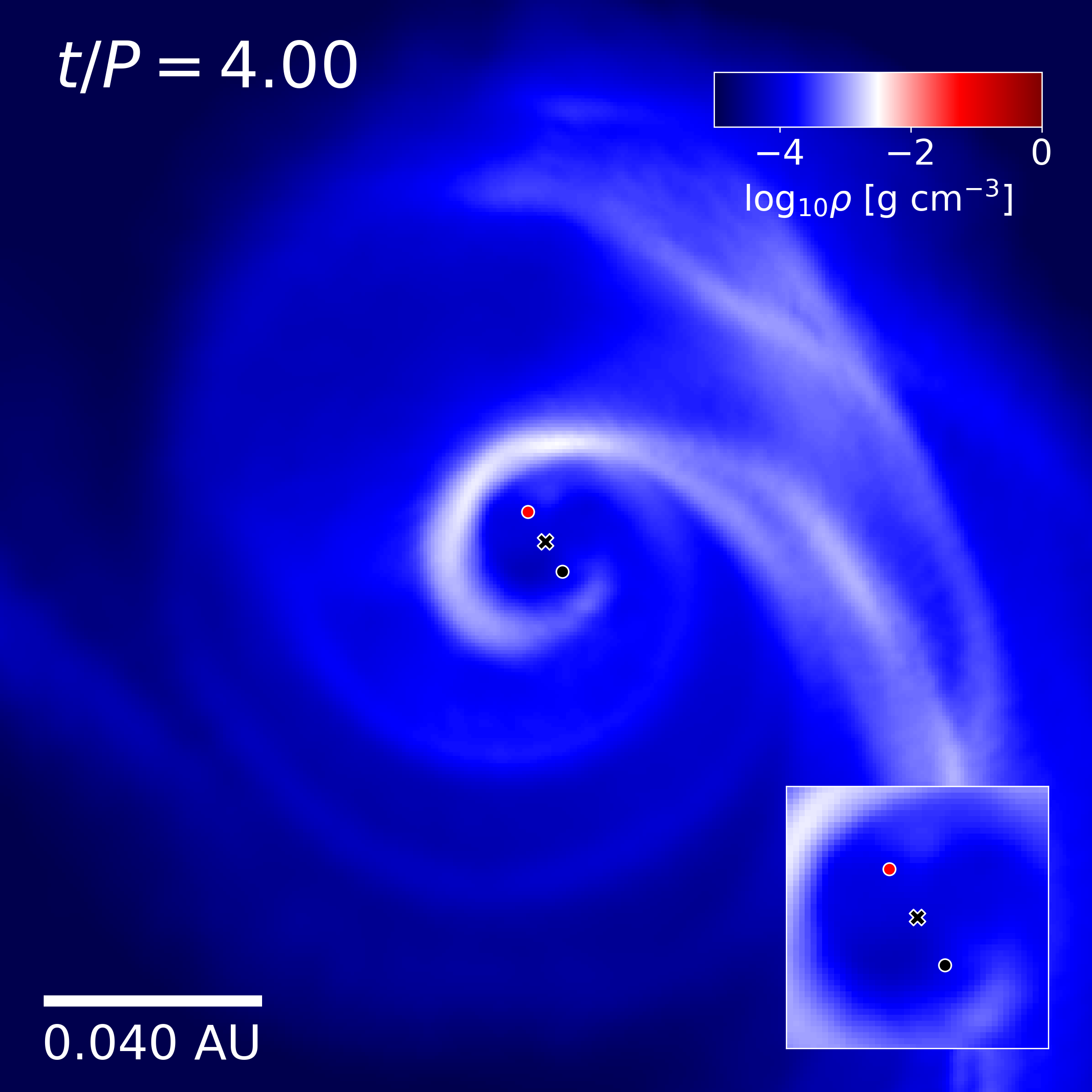}
\includegraphics[width=4.3cm]{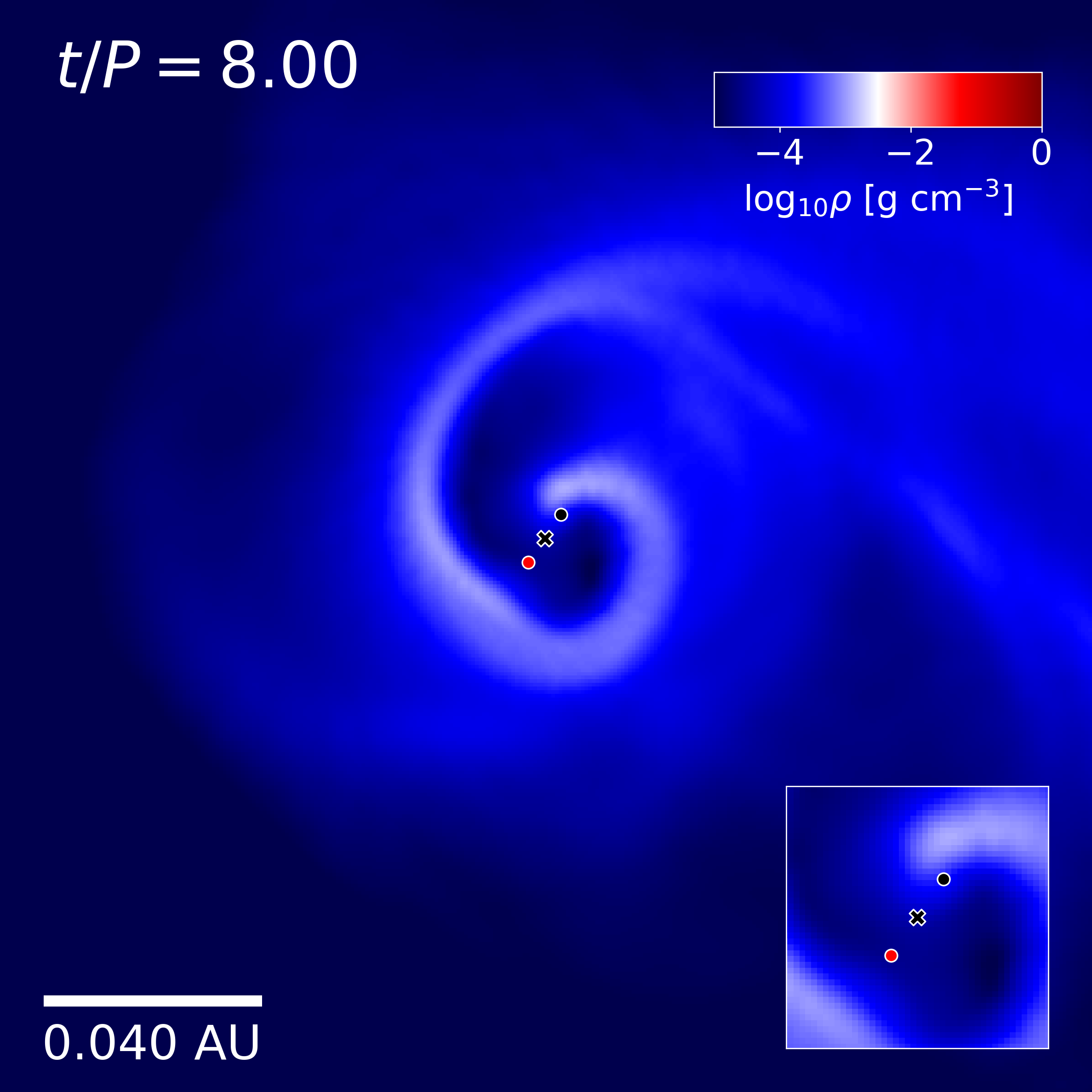}
	\caption{The density distribution in the binary orbital plane in two models (\textit{bottom}: $\alpha4-i180$ and \textit{top}: $Fiducial$) at a few different times. The time is measured since $t=t_{\rm arr}$ in unit of the binary period $P$. The color bar shows the logarithmic density in ${\rm g}/{\rm cm}^{3}$. The insets focus on the regions near the center of mass (the cross) of the binary (the black and red dots). In the \textit{top} panels, the star approaches on a retrograde orbit from the bottom-right corner (the blue dotted curve in Figure~\ref{fig:IC}) and is disrupted by the BH in black. The debris starts to significantly interact with the binary in red only after an eccentric disk forms due to stream-stream intersection. 
	In the \textit{bottom} panels, the star approaches the BH in red on a prograde orbit from the bottom-left corner (the green dotted curve in Figure~\ref{fig:IC}) and is disrupted by the BH in red. The debris interacts with the bystander BH (the black dot) promptly and the debris flow becomes highly irregular. }	\label{fig:example} 
\end{figure*}

\section{Results}\label{sec:result}

\subsection{Overview of the encounter dynamics}\label{subsec:overview}

We split our models into two categories based on the trajectory of the debris stream after the first encounter and on how quickly the debris interacts with the bystander BH (the non-disrupting BH): \textit{regular debris flow} and \textit{non-regular debris flow}. We describe the stream evolution for each case using the two examples illustrated in Figure~\ref{fig:example} where we show the density distribution in the binary orbital plane at a few different times, measured since $t_{\rm arr}$. Note that we indicate which case each model corresponds to in Table~\ref{tab:orbitchange}. 

\begin{enumerate}
    \item \textit{Regular debris flow} (\textit{top} panels of Figure~\ref{fig:example}): the star is disrupted outside the binary (e.g., at $t/P=0$). The debris travels outward until bound mass returns (e.g., at $t/P=0.5)$. The shape and trajectory of the debris look very much like regular TDEs by single black holes. The early returning most-bound debris intersects with the newly-coming debris near apocenter ($t/P=4$) and forms an eccentric disk either around the binary (e.g., Model $\alpha4-ret.$) or the disruptor BH (e.g., Model $a0.1-retro.$). However, unlike a TDE by a single BH, the eccentric stream is continuously perturbed by the time-dependent potential of the binary on a time scale comparable to the binary orbital period. For some cases with an eccentric accretion flow around the disruptor BH, the stream overflows to the other BH (e.g., Models $a0.03-pro.$ and $a0.1-pro.$). Because of this quasi-periodic perturbation, the accretion rate of the BHs for this case reveals modulations (\S\ref{subsec:accretion}). The encounter can harden or widen the binary, but its immediate impact on the binary's orbit tends to be smaller than for the \textit{non-regular debris flow} case. This case occurs when the star is disrupted at a relatively large distance from the binary (hence no significant perturbations of the orbit by the disruptor BH) or when the star is on a hyperbolic orbit around the disruptor BH. All the models where the star loses its mass partially at the first encounter (e.g., Models $\alpha1-pro.$ and $\alpha1-retro.$) fall into this category. 

    \item \textit{Non-regular debris flow} (\textit{bottom} panels of Figure~\ref{fig:example}): the star's orbit deviates significantly from the original orbit near the disruptor BH. The star is then disrupted by the disruptor BH's tidal gravity as it passes through the new pericenter  (e.g., Model $Fiducial$) or by a nearly direct collision with the disruptor BH (e.g., Models $q0.3S-pro.$ and $q0.3S-retro.$). For the case where the star is tidally disrupted, since the star's velocity is faster than the disruptor BH and the pericenter distance of the new orbit around the disruptor BH is much smaller than the binary semimajor axis, the disrupted star quickly passes through the pericenter and heads back towards the direction of initial approach. This quick pericenter passage, followed by a turn-around, results in two important outcomes. First, change of motion into the opposite direction gives a forward (backward) momentum kick to the disruptor BH for prograde (retrograde) encounters, which results in a sudden widening (hardening) of the binary (see \S\ref{subsec:binary_orbit}). Second, the debris promptly interacts with the bystander BH which is on the way out of the debris (e.g., at $t/P=0$). This subsequent interaction significantly perturbs the entire debris streams (e.g., at $t/P=0.5$). As a consequence, the subsequent interactions between the debris and binary are very violent and irregular. 
    
    We find two head-on collision cases when the less massive BH in the unequal-mass binary is the disruptor BH (Models $q0.3S-pro.$ and $q0.3S-retro$). The star's trajectory near the binary is significantly deflected towards the bystander BH (the more massive one), which takes the star on an almost head-on collision course to the less massive one. Like the \textit{non-regular debris flow} case involving a TDE, the gas promptly interacts with the bystander BH, resulting in an irregular stream evolution.
    
\end{enumerate}

\citet{Lopez2019} considered three dynamical scenarios for the star-binary close encounters while they were investigating the impact of TDEs on the BH spin using hydrodynamics simulations. Based on their characterization of the encounter properties and debris evolution, their circumbinary scenario corresponds to our \textit{regular debris flow}. Their overflow scenario describes a subset of the models corresponding to \textit{non-regular debris flow} (e.g., Models $a0.03-pro.$ and $a0.1-pro.$). There are no cases in our simulations that share the same features with their single case where the debris only interacts with the disruptor BH. This is likely because the two studies consider the semimajor axes in very different ranges: $0.2\AU\leq a\leq 2\AU$ in their work and $0.01\AU\leq a\leq 0.1\AU$ in this work. The debris-binary interactions are likely more prompt and violent in our simulations because of the smaller semimajor axes and, hence, stronger perturbing forces from the bystander BH, which makes it hard for the debris to interact only with one BH.

\begin{figure}
\includegraphics[width=8.6cm]{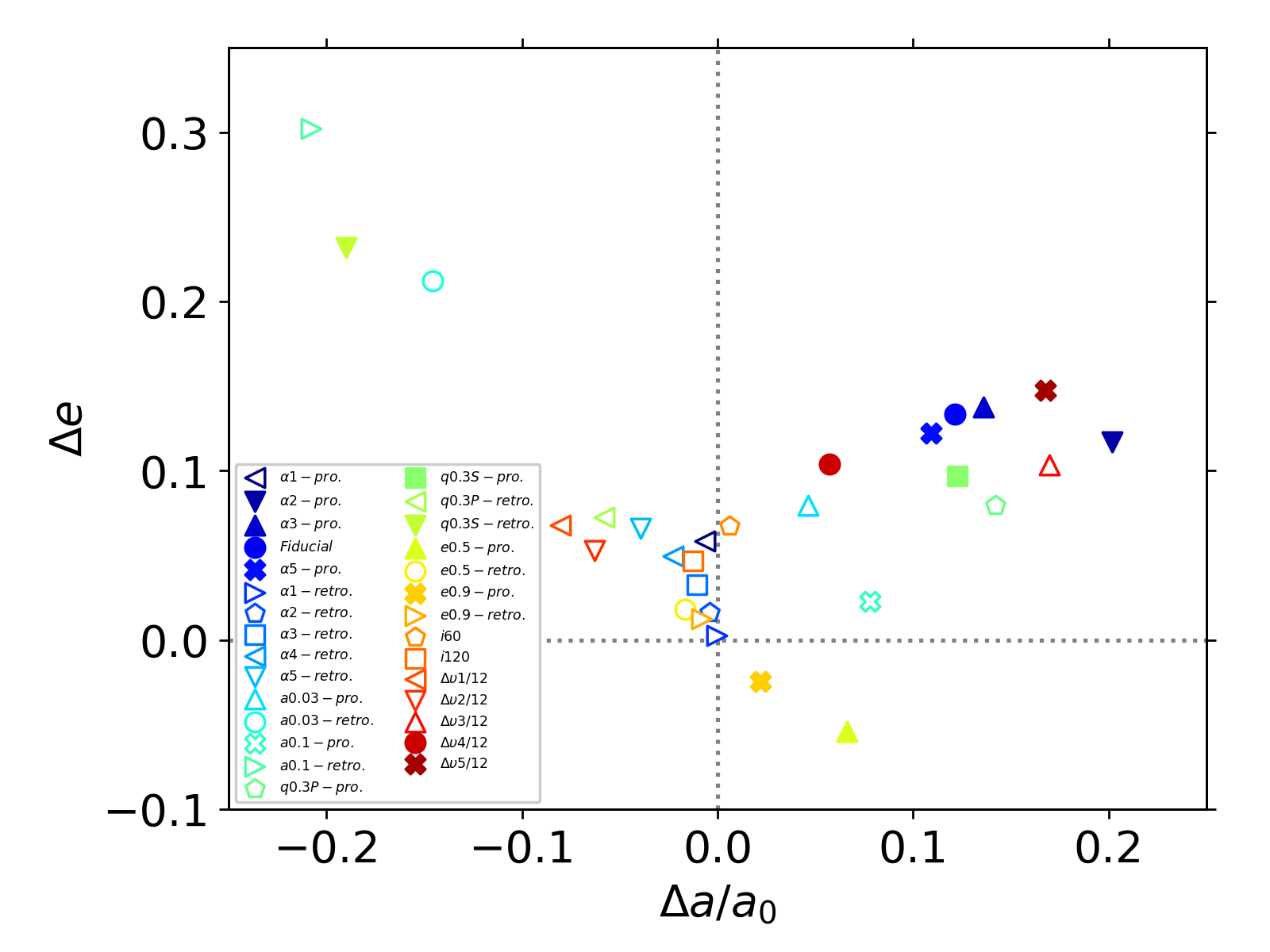}
	\caption{The change in the binary orbital elements in the \textit{non-regular debris flow} (solid markers) and \textit{regular debris flow} (hollow markers) cases. $\Delta a/a_{0}$ is the fractional change of $a$ relative to the initial value $a_{0}$ and $\Delta e$ the change of $e$ after the encounter relative to the initial value. }	\label{fig:deltae_deltaa} 
\end{figure}

\subsection{Binary orbit after close encounter}\label{subsec:binary_orbit}

We present the fractional change in $a$ relative to the initial value, $\Delta a/a_{\rm 0}=(a-a_{0})/a_{0}$, and the absolute change in $e$ from the initial value, $\Delta e= e- e_{0}$, due to the first close encounter in Figure~\ref{fig:deltae_deltaa}. Note that the changes in the orbital parameters are mostly due to the first encounter, not the interactions between the debris gas and the binary. There are a few noticeable trends. 

First,  prograde encounters (solid markers) tend to make the binary less compact ($\Delta a>0$) whereas retrograde encounters (hollow markers) tighten the binary ($\Delta a<0$). %, with the three outliers (Models $\alpha1-prog.$, $\alpha \Delta\upsilon1/12$ and $\Delta\upsilon2/12$). This can be understood as follows. 
 In the \textit{non-regular debris flow} case, the disruption looks like an ordinary TDE in the frame of the disruptor BH until the debris starts to interact with the other BH. Because the incoming motion of the star and outgoing motion of the debris (also in the disruptor BH's frame) are in the opposite directions, the BH receives a momentum kick. For prograde (retrograde) orbits, the disruptor BH gains (loses) momentum, which leads to widening (hardening) of the binary. In the \textit{regular debris flow} case, whether the binary hardens or widens may sensitively depend on the relative velocity between the disruptor BH and star and the binary phase. However, the encounters more likely make the binary harder because the velocity of the star is faster than the disruptor BH (in the CoM frame of the binary). In both prograde and retrograde encounters, the fractional change $|\Delta a/a_{0}|\lesssim 0.2$, but in-plane encounters have a bigger impact than off-plane encounters (Models $i60$ and $i120$). 
 
Second, for initially circular binaries, both prograde and retrograde encounters make the binary eccentric (roughly by  $\lesssim 15\%$), which is not surprising. However, for initially eccentric binaries, prograde and retrograde encounters change the eccentricity in the opposite sense: prograde encounters ($e0.5-pro.$ and $e0.9-pro.$) circularize the binary ($\Delta e\gtrsim -0.05)$ whereas for retrograde encounters ($e0.5-retro.$ and $e0.9-retro.$), the binary becomes more eccentric after the disruption. Interestingly, $\Delta e$ for such encounters is especially large, i.e., $\simeq 0.2-0.3$. 

Third, for unequal-mass binaries, the orbit elements are affected more when the disruptor BH is the less massive one. Both $|\Delta a|/a_{0}$ and $\Delta e$ are $\sim0.1-0.2$ for the case where the star is disrupted by the secondary BH. However, for the other case, $|\Delta a|/a_{0}\lesssim 0.1$ and   $\Delta e\simeq 0.08-0.09$.

It is interesting to note that the different outcomes arise simply because of different phase angles at the encounter (Models starting with $\Delta\upsilon$). The star is partially disrupted at the first close encounter in Models $\Delta\upsilon2/12$ and $\Delta\upsilon3/12$ whereas it is fully disrupted in the rest of the models with different phase angles including the fiducial case. Furthermore, the stream evolution is quite different: three \textit{regular debris flow} cases and three \textit{non-regular debris flow} cases. This implies that even when the initial orbital parameters of the incoming star are the same, not all close encounters result in TDEs simply because of different binary phases at the encounter. For this reason, the phase angle could be an important determining factor for the fate of the star at the first close encounter as well as the subsequent stream evolution.

We summarize the values of $a$, $e$ and $t_{\rm GW}$ before and after the encounter in all our models in Table~\ref{tab:orbitchange}.

\begin{figure}
\includegraphics[width=8.6cm]{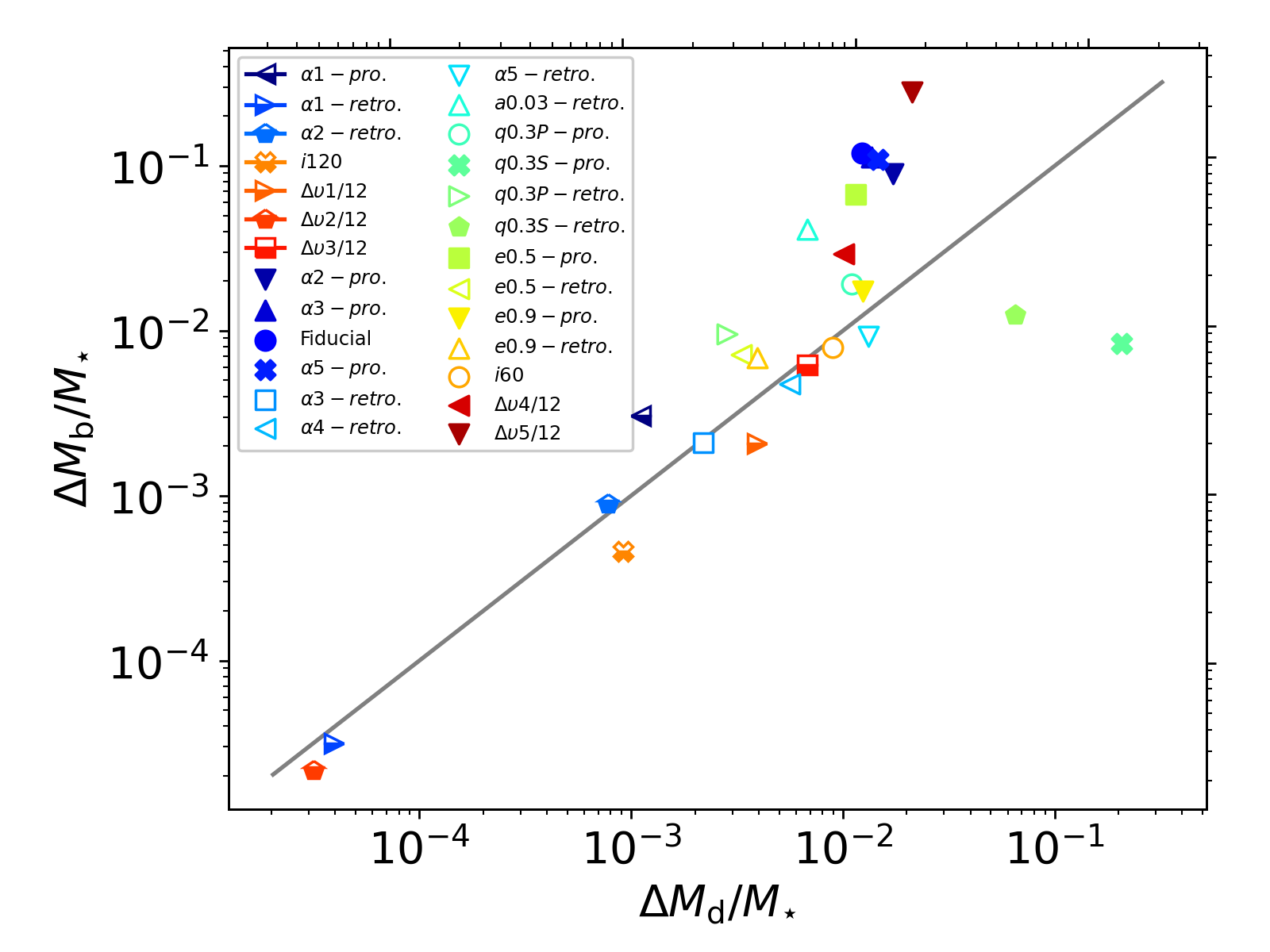}
	\caption{The fractional accreted mass of the disruptor BH, $\Delta M_{\rm d}/M_{\star}$, and the bystander BH, $\Delta M_{\rm b}/M_{\star}$, measured at $t=10P$ after the first encounter. The models that end up with a partial disruption are depicted with half-filled markers. Among full disruptions, we distinguish the \textit{regular debris flow} and the \textit{non-regular debris flow} using hollow and solid markers, respectively.}	\label{fig:macc} 
\end{figure}

\begin{figure*}
\includegraphics[width=8.6cm]{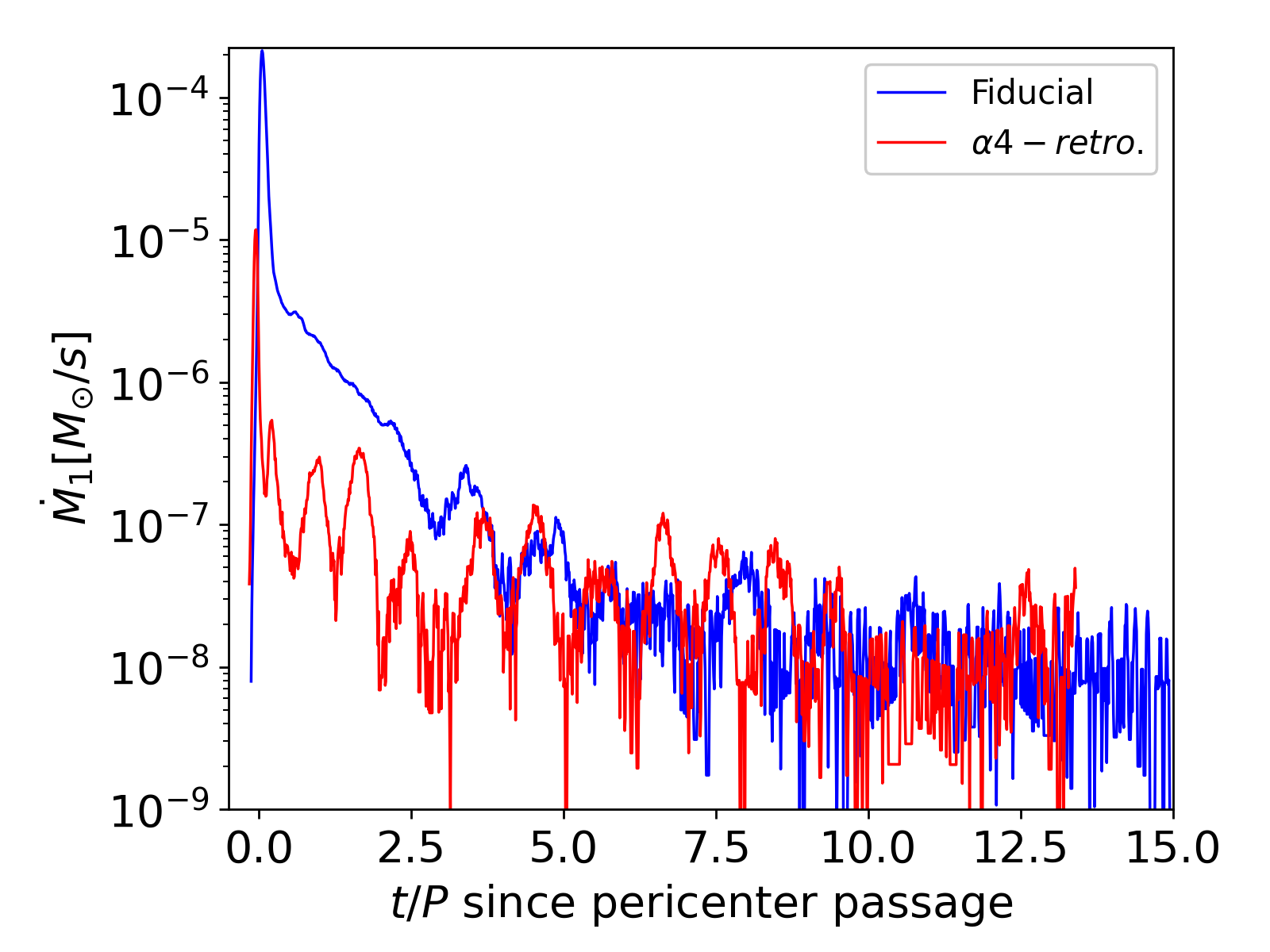}
\includegraphics[width=8.6cm]{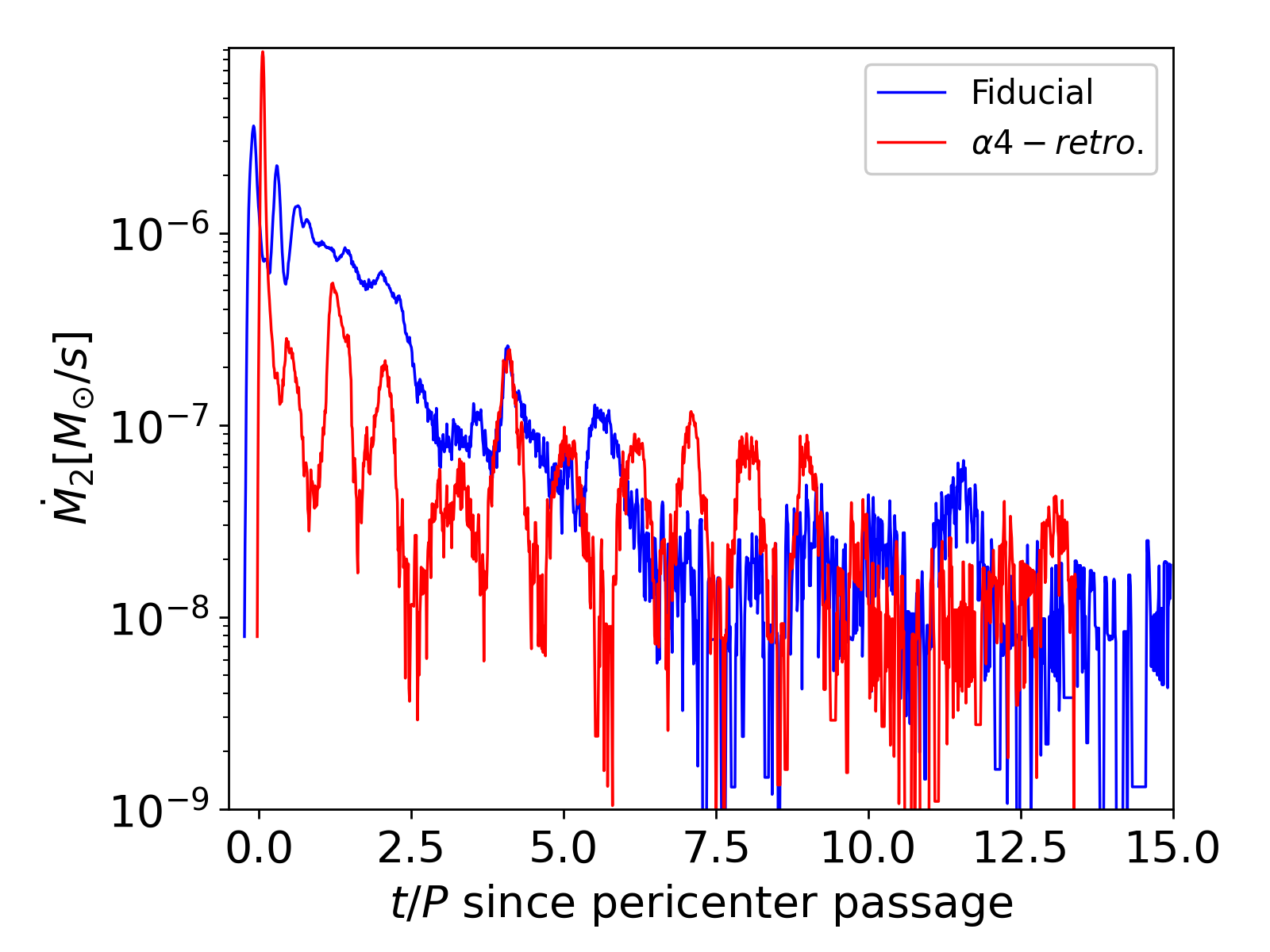}
	\caption{The accretion rate of the primary (\textit{left}) and secondary (\textit{right}) black holes in Models $fiducial$ (blue) and $\alpha4-i180$ (red), which are example cases for the \textit{non-regular debris flow} and \textit{regular debris flow} cases, respectively. }	\label{fig:acc} 
\end{figure*}

\subsection{Mass accretion}\label{subsec:accretion}

In our simulations, BHs grow in mass via accretion. Figure~\ref{fig:macc} shows the fractional accreted mass of the disruptor BH, $\Delta M_{\rm d}$, and the bystander BH, $\Delta M_{\rm b}$, measured at $t=10P$ since the first encounter. It is not surprising that the total fractional accreted mass $(\Delta M_{\rm d}+ \Delta M_{\rm b})/M_{\star}$
in partial disruptions (which is $\lesssim 10^{-1}$) is smaller than that in full disruptions. For full disruptions, we find quite noticeable differences between the \textit{regular debris flow} and the \textit{non-regular debris flow} cases. which have a fractional accreted mass  $\gtrsim 10^{-1}$. Generally, the accreted masses ($\Delta M_{\rm d}$ and $\Delta M_{\rm b}$) in the \textit{regular debris flow} are smaller than those in the \textit{non-regular debris flow} case.
Furthermore, $\Delta M_{\rm d}$ and $\Delta M_{\rm b}$ are quite comparable in the \textit{regular debris flow} whereas $\Delta M_{\rm b}$ is typically greater than $\Delta M_{\rm d}$. These trends in fact reflect the nature of the binary-stream interaction. In the \textit{non-regular debris flow} case, right after disruption the debris undergoes violent (almost head-on like) interactions with the bystander BH, which results in a mass accretion burst. Because the subsequent mass accretion onto the two BHs is quite symmetric, the accreted mass of the bystander BH is greater than that of the disruptor BH. On the other hand, 
in the \textit{regular debris flow} case, there is no accretion burst, but rather periodic and symmetric mass accretion onto the two BHs. This is because of repeated interactions of the binaries with the debris stream that returns and orbits close to the binary. Each element of the debris upon disruption moves on a ballistic orbit around the binary and eventually the bound mass returns to the binary. The debris passing through the pericenter, which is similar to the that of the original stellar orbit, is deflected towards the binary, which naturally leads to  interactions of the debris with the black hole orbiting near the stream. This is illustrated in the two top right panels in Figure~\ref{fig:example}.

Note that there are two outliers (Models $q0.3S-pro.$ and $q0.3S-retro.$) with $\Delta M_{\rm d}>\Delta M_{\rm b}$. In the two cases, the secondary BH, which is the disruptor BH, goes through a head-on collision with the star at the first encounter. During the collision, the secondary BH accretes a significant amount of mass. Like other  \textit{non-regular stream cases}, the subsequent accretion onto the two BHs is more or less symmetric. 

The different mass accretion episodes between \textit{regular debris flows} and \textit{non-regular debris flows} can be seen clearly from the accretion rate. We show the accretion rate for two representative cases (Models $Fiducial$ and $\alpha4-retro.$) in Figure~\ref{fig:acc}. In both cases, the secondary BH (\textit{right} panel) is the disruptor BH. We calculate the accretion rate by dividing the accreted mass of each BH by the time between two very finely sampled adjacent output data. In the \textit{non-regular debris flow} case (Model $Fiducial$), 
the peak accretion rate of the bystander BH (\textit{left} panel) is higher due to the violent interaction right after disruption. But the post-peak rates are quite similar. On the other hand, in the \textit{regular debris flow} case (Model $\alpha4-retro.$), the accretion rates of the two BHs are very similar. Remarkably, the periodic binary-stream interaction results in quasi-periodic modulations of the accretion rate on a timescale comparable to $\simeq P$. This type of behavior is clearly more regular and found more frequently in the \textit{regular debris flow} cases. All of the models that fall into this class show such modulations at least up to $\simeq 5P$ after the close encounter.   

In fact, the modulation of the accretion rate could serve as a characteristic feature that distinguishes TDEs by BBHs from those by single stellar-mass BHs. Given that these features originate from the gravitational capture of gas particles, followed by accretion in periodic interactions between debris streams and the BHs, even if we simulated the region inside the accretion radius, these features would be robust although the accretion rate itself would be lower than that found in our simulations. 

Another important feature of BBH-driven TDEs which is different from single BH-driven TDEs is that the fallback rate and the accretion rate are rather decoupled. For TDEs by a single stellar-mass black hole, the most bound debris that has returned to the BH and passed through pericenter forms already a weakly eccentric to almost circular disk around the BH ($1-e\simeq 0.74(\Mbh/20\Msol)^{-1/3}(M_{\star}/1\Msol)^{1/3}$). Thus the debris can fully circularize quite rapidly. Under such conditions, the viscous time scale $t_{\rm vis}$ is greater than the fallback time scale $t_{\rm fb}$  for typical parameters by a factor of a few, i.e.,
\begin{align}
    \frac{t_{\rm fb}}{t_{\rm vis}}\sim0.3\left(\frac{\alpha}{0.1}\right)(h/r)^{2}\sqrt{\frac{\Mbh}{20\mstar}}.
\end{align}
%\begin{align}
%    13 \lesssim \left(\frac{\alpha}{0.1}\right)(h/r)^{2}\sqrt{\frac{\Mbh}{\mstar}}\;.
%\end{align}
Here, $h/r$ is the disk aspect ratio and $\alpha$ the viscous constant. This means the viscous dissipation rate would primarily determine the accretion rate. However, a correlation between the fallback rate and the accretion rate may be still expected. On the other hand, for TDEs by stellar-mass BBHs,  no correlation is expected because there is no regular mass fallback (\textit{non-regular debris flow}), or  accretion occurs via repeated interactions between the binary and the stream (\textit{regular debris flow}). As a result, the accretion rate cannot be described by a characteristic powerlaw slope as in the case of TDEs around a single BH.

\begin{figure}
\includegraphics[width=8.6cm]{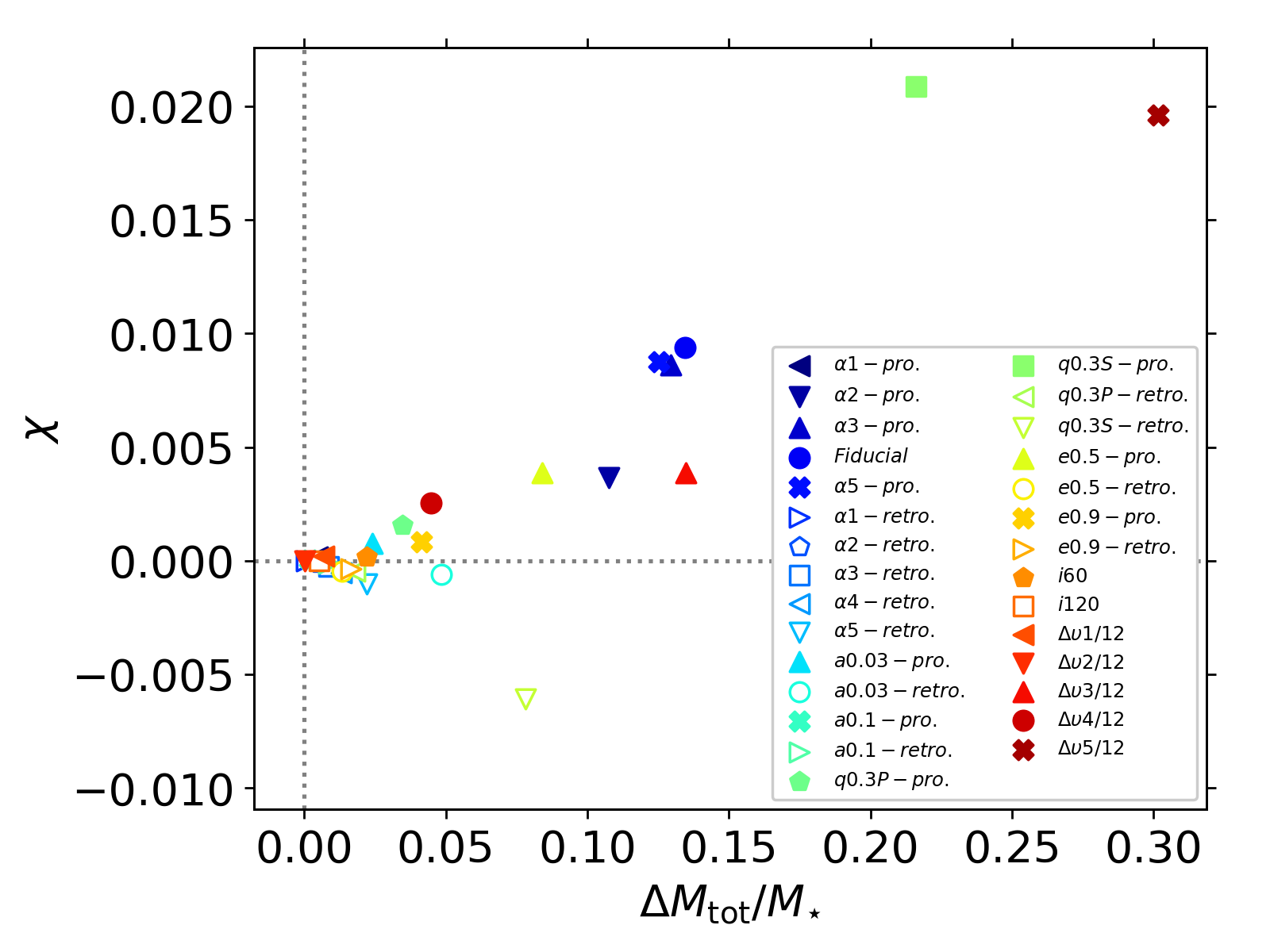}
	\caption{The effective spin parameter $\chi$ (Equation~\ref{eq:chi}) measured at $t\simeq 10P$ for $a\leq 0.03\AU$ and $t\simeq P$ for $a=0.1\AU$ since the first encounter, as a function of the ratio of the total accreted mass $\Delta M_{\rm tot}=(\Delta \Mbhp+\Delta \Mbhs)$ to $M_{\star}$. We differentiate prograde encounters from retrograde encounters by using different marker filling style: prograde (solid) and retrograde (hollow).   }	\label{fig:chi} 
\end{figure}

\begin{figure*}
\includegraphics[width=8.6cm]{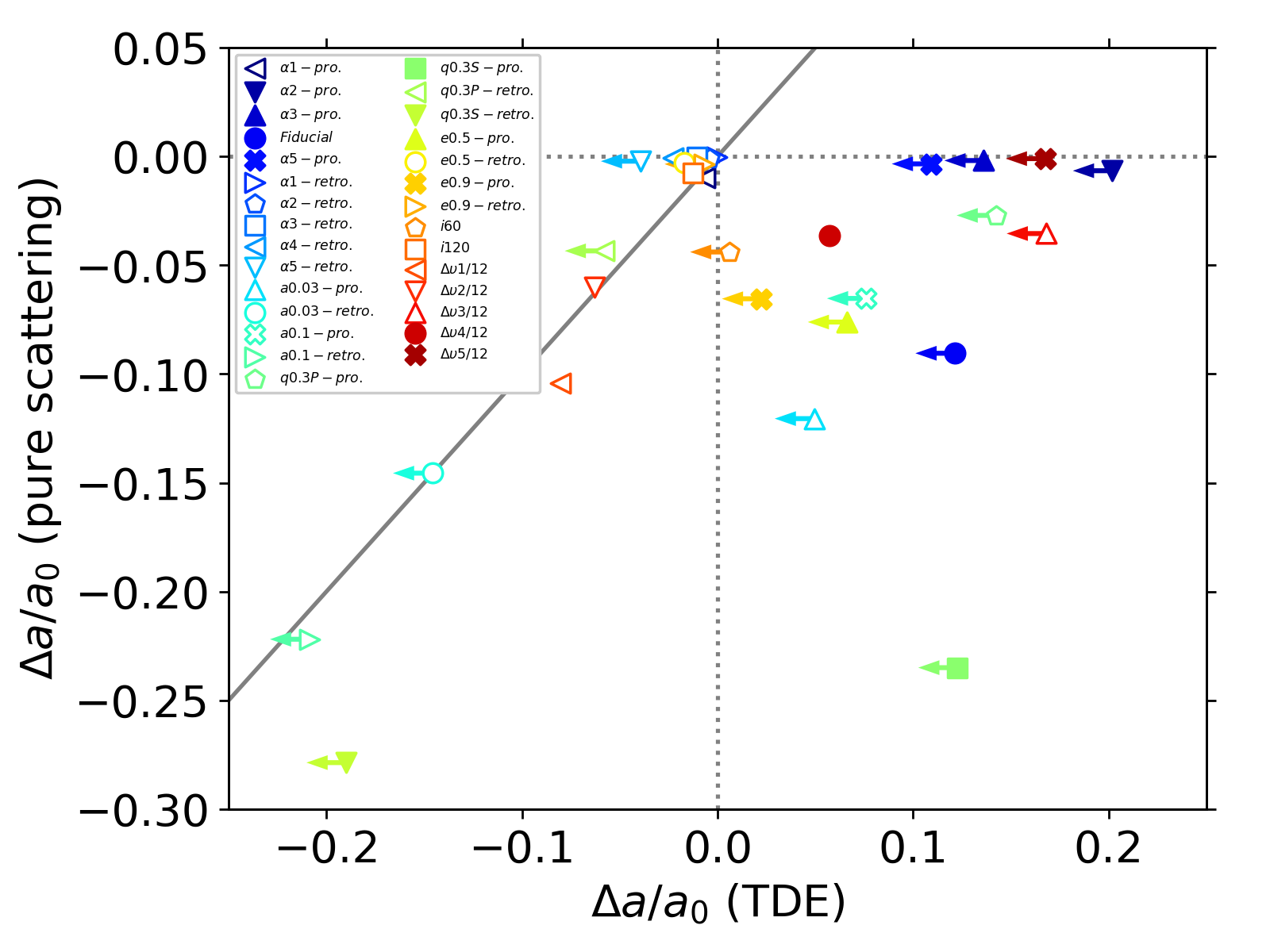}
\includegraphics[width=8.6cm]{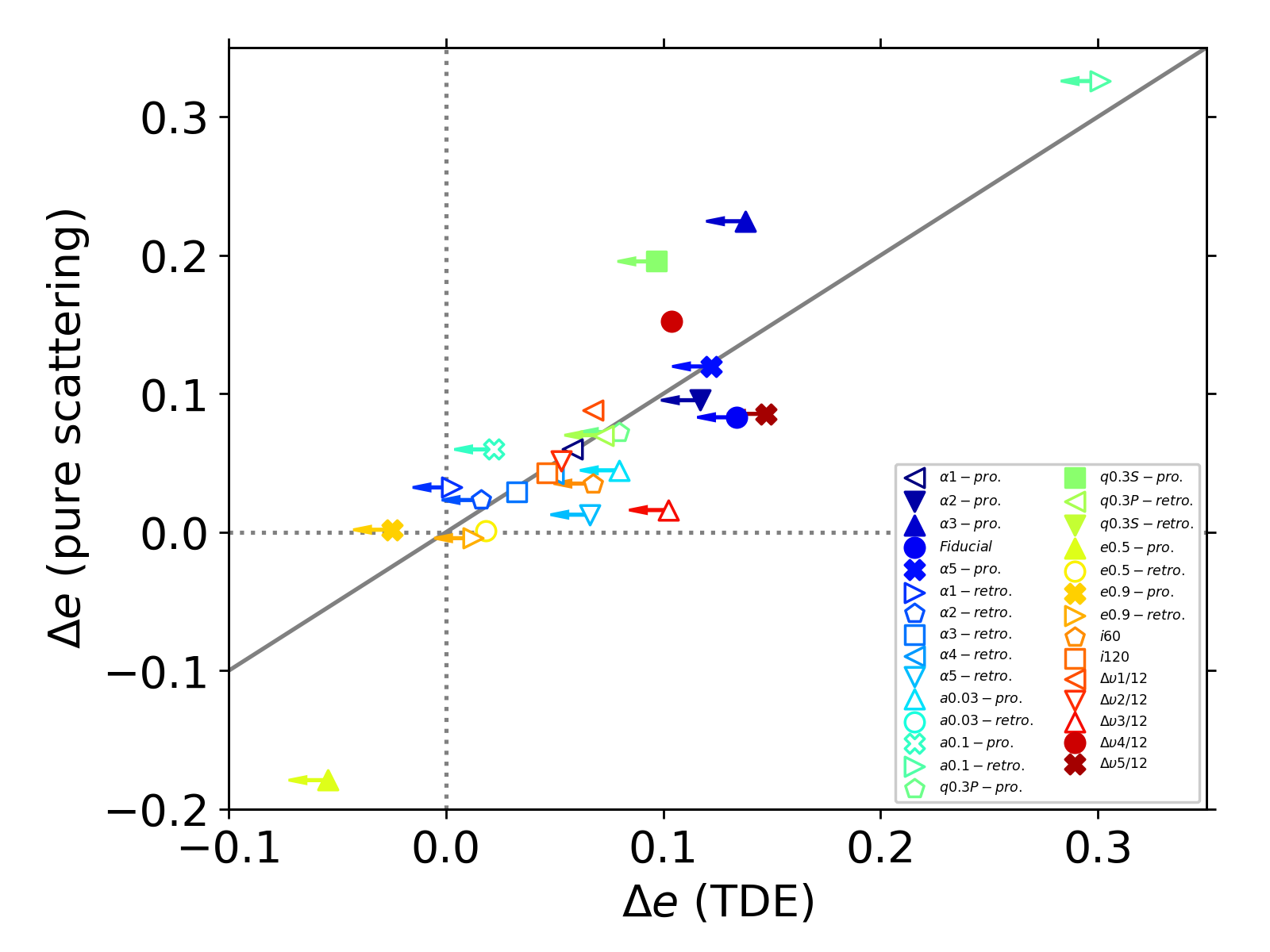}
	\caption{Comparison between $\Delta a/a_{0}$ (\textit{left}) and $\Delta e$ (\textit{right}) for TDEs and pure scatterings, measured at the end of the simulations. We use the same initial conditions for each TDE case to perform pure scattering experiments. The solid (hollow) markers indicate the case that falls into the \textit{non-regular debris flow} (\textit{regular debris flow}) case. The markers with (without) a right-pointing arrow refer to the case where the binary undergoes multiple chaotic (single) encounters.  }	\label{fig:purescattering} 
\end{figure*}

\begin{figure}
\includegraphics[width=8.6cm]{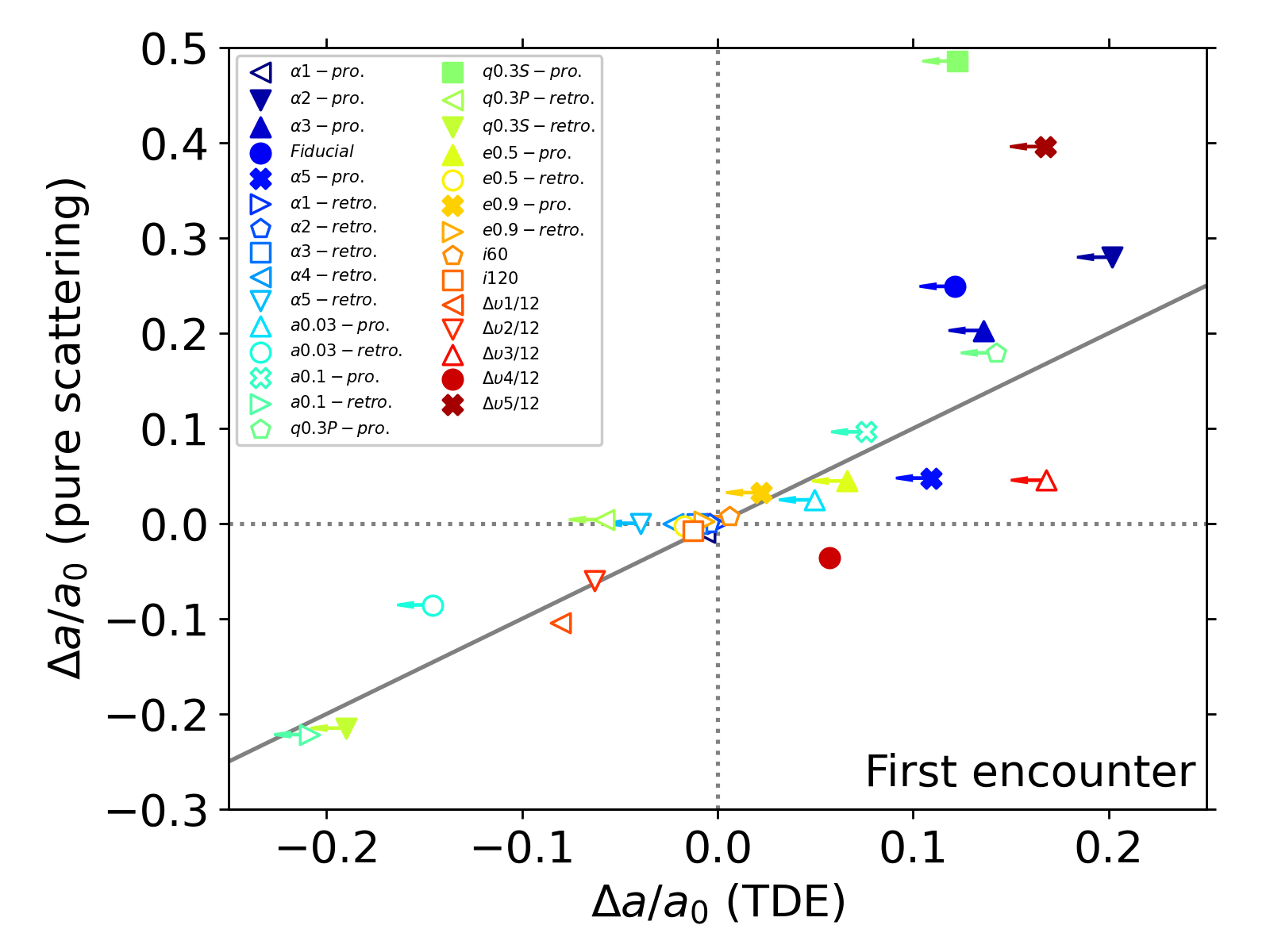}
	\caption{Same as in the \textit{left} panel of Figure~\ref{fig:purescattering}, but $\Delta a/a_{0}$ for pure scattering is measured after the first close encounter. }	\label{fig:purescattering_first} 
\end{figure}

\subsection{Angular momentum accretion and spin}

The accretion of gas can change the BH spin. We compute the change in BH spin due to mass accretion, following the formalism by \citet{Brown+2000} with the assumption that the angular momentum vector of the material crossing the accretion radius is parallel to that at the last stable orbit. More precisely, when the BH with spin angular momentum $\textbf{J}$ accretes an amount of material with mass $M_{\rm acc}$ and angular momentum $\textbf{J}_{\rm acc}$ at a given time step, its new angular momentum $\textbf{J}'$ is,
\begin{align}
   \textbf{J}' = \textbf{J} + \textbf{J}_{\rm acc}.
\end{align}
Here, $\textbf{J}_{\rm acc}$ is the angular momentum vector of the material with mass $M_{\rm acc}$ at the last stable orbit, $\textbf{J}_{\rm acc}=J_{\rm acc} (\tilde{J}/|\tilde{J}|)$.  Its magnitude $J_{\rm acc}$ is only a function of $M_{\rm acc}$ (here it is calculated using Equations 11-13 in \citealt{Perna+2018}), and $\tilde{J}$ is the angular momentum vector of the material crossing the accretion radius. Then the effective spin parameter $\chi$ is defined as
\begin{align}\label{eq:chi}
   \chi = \frac{\Mbhp \textbf{a}_{1} + \Mbhs\textbf{a}_{2}}{\Mbhp+\Mbhs} \cdot \textbf{L},
\end{align}
where $\textbf{a}_{i}=\textbf{J}'_{i}c/M_{\rm BH,i}^{2}G$ ($i=1,2$) and $\textbf{L}$ is the unit orbital angular momentum. Assuming instantaneous in-plane accretion of gas with mass $0.5\Msol$, the initially non-rotating $M_{\rm BH}=20\Msol$ ($6\Msol$) BH can have the maximum value of $\chi\simeq 0.09 (0.3)$. Note that in reality, because only a fraction of the material crossing the accretion radius would actually accrete onto the BH, our estimate for $\chi$ is likely to be an upper limit. 

Figure~\ref{fig:chi} shows the effective spin parameter $\chi$ as a function of the total accreted mass $\Delta M_{\rm tot}=(\Delta M_{\rm d}+\Delta M_{\rm b})$, normalized by $M_{\star}$, measured at $t\simeq 10P$ for $a\leq 0.03\AU$ and $t\simeq P$ for $a=0.1\AU$ since the first encounter. As expected,  $|\chi|$ is linearly proportional to the accreted mass: $\chi\simeq 0.01 (0.02)$ for $M_{\rm tot}/M_{\star}\simeq 0.13 (0.3)$. $\chi$ tends to be larger for the \textit{non-regular debris flow} cases because of the larger accreted mass (Figure~\ref{fig:macc}).

\begin{figure}
\includegraphics[width=8.6cm]{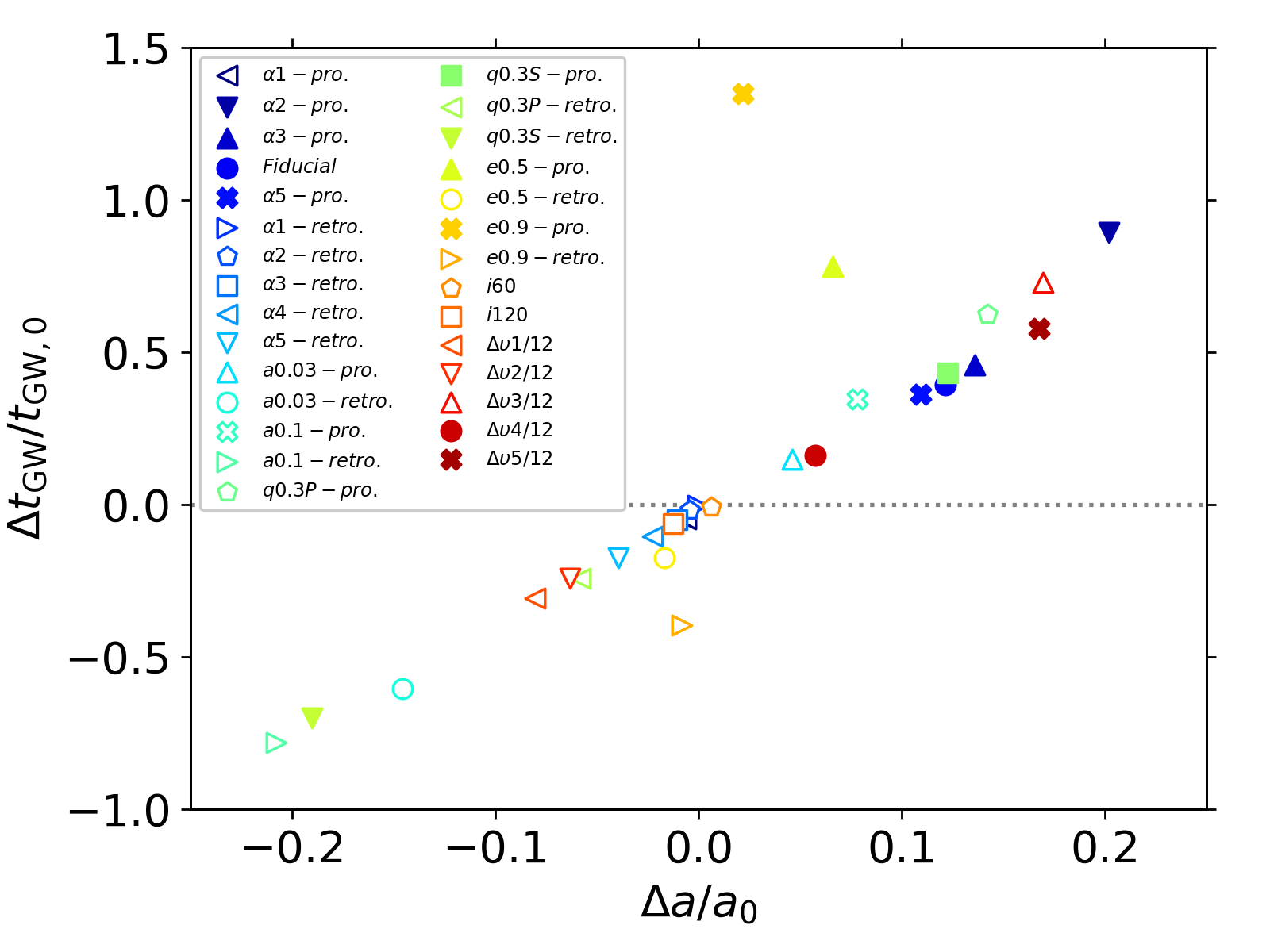}
	\caption{The fractional change in the GW-driven merger time scale $t_{\rm gw}$, relative to the $t_{\rm gw}$ for the original binary, for all our models. The solid (hollow) markers indicate the \textit{non-regular debris flow} (\textit{regular debris flow}) cases. }	\label{fig:frac_tgw} 
\end{figure}

\begin{figure*}
\includegraphics[width=17.6cm]{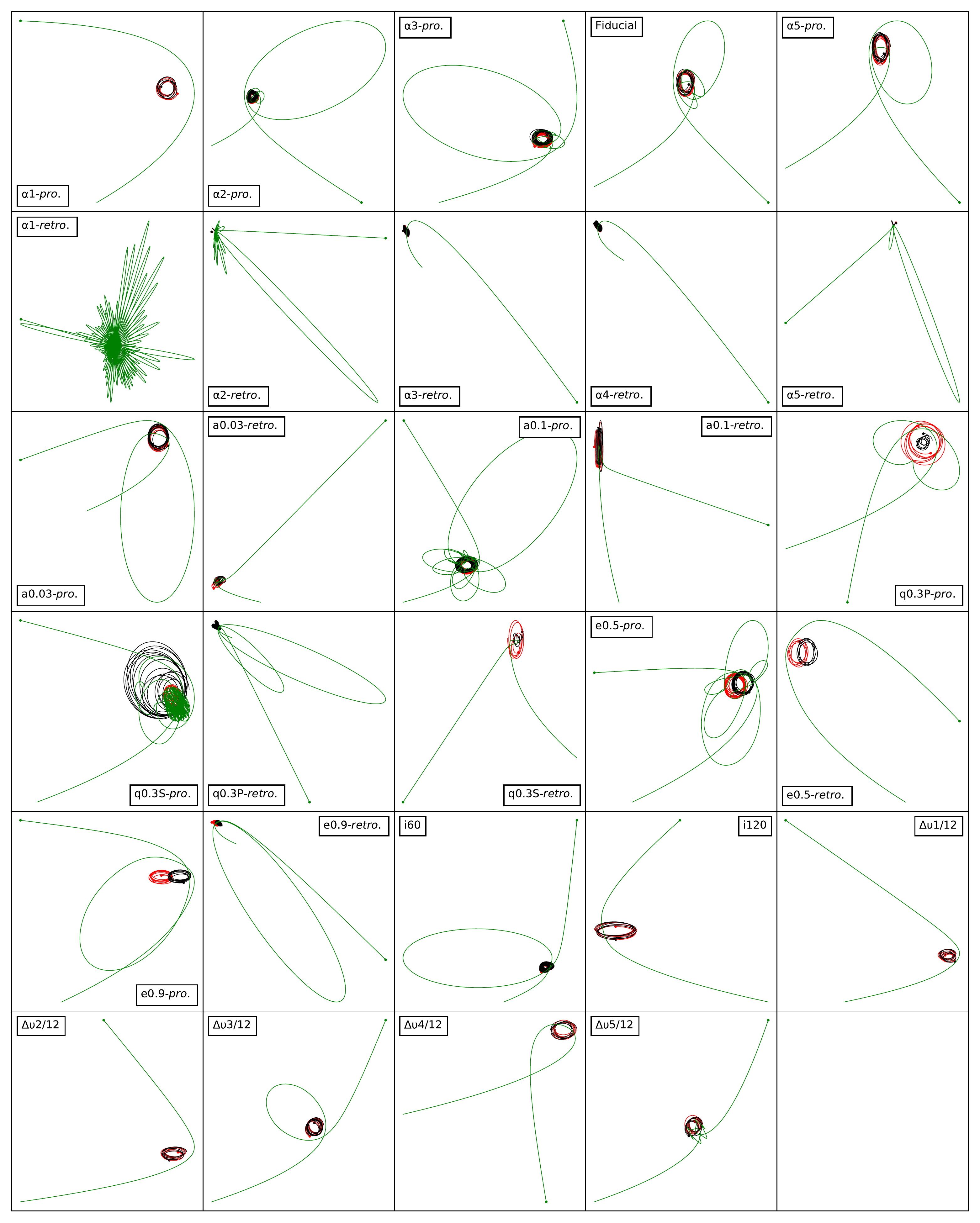}
	\caption{The trajectory of the BBH (black and red lines) and the star (green lines) projected on the initial binary orbital plane in the pure scattering experiments. The final outcome  in all our models is  binary + ejected single. The final position of the ejected star is marked by the green circle located at one end of the green line.  }	\label{fig:trajectory_purescattering} 
\end{figure*}

\section{Astrophysical Implications}\label{fig:implication}
\subsection{Comparison with pure scattering}

The theory of a few-body dynamics is built upon analytic and numerical simulations in which stellar objects are point particles \citep{Hut1983,Hut1993,Heggie1996,Hamers2019a,Hamers2019b}, although some previous studies include tidal effects in their calculations \citep{Fabian1975, Press1977,Lee1986,McMillan1986, Samsing2017,Samsing2018}. However, it is possible that stars can be tidally disrupted during chaotic or secular interactions with black holes, which can have a different impact on the dynamics. 

In order to compare the impacts of TDEs with those of pure scatterings, we perform three-body scattering experiments using the high precision few-body code {\sc SpaceHub} \citep{spacehub}. The two sets of simulations share the identical initial conditions for each model, but the key difference is whether the star is assumed to be a point particle or an object subject to hydrodynamic and tidal interactions. We run the three-body simulations until the three objects establish a stable configuration: 1) binary + ejected single, 2) ejected three singles and 3) a stable triple. However, because the initial total energy in all our models is negative, the formation of ejected three singles is not allowed. The only final outcome is binary + ejected single, and no stable triple forms. We show the trajectory of the binary and star for all our models in Figure~\ref{fig:trajectory_purescattering}.

We find that the change in the binary's semimajor axis due to TDEs can be significantly different from that due to pure scatterings. More specifically, TDEs can lead to both hardening and widening of our binaries whereas the binaries get only hardened by pure scatterings. This is illustrated in the \textit{left} panel of Figure~\ref{fig:purescattering} where we compare the final $\Delta a/a_{0}$ for all the models between the hydrodynamics and pure-scattering simulations. It is not surprising that the binaries become more compact in pure scattering cases because the generally expected outcome of scatterings between a hard binary and a star is the same binary that becomes more bound and an ejected star (negative $\Delta a/a_{0}$ (pure scattering) in Figure~\ref{fig:purescattering}).

In particular, a significant contrast in  $\Delta a/ a_{0}$ between TDEs and pure scatterings arises in the case where the binary becomes wider after a TDE (solid markers, mostly prograde encounters). This happens to be the case where the star undergoes multiple encounters with the binary when the star and the binary are assumed to be point particles (see Figure~\ref{fig:trajectory_purescattering}). The three-body scatterings in the pure scattering regime can be either non-resonant (one encounter before ejection) or resonant (multi-encounters before ejection). We find 8 models (21 models) in the resonance (non-resonance) scattering regime. Because the binary was already hard, the final outcome tends to be a more hardened one. However, interestingly, we find that the first encounter in the resonance scattering regime (both TDEs and pure scatterings) tends to make the binary softer. TDEs effectively act to let the binary be only affected by the very first encounter, having the similar impact on the binary orbit as the pure scattering, and not undergo subsequent chaotic encounters. To demonstrate this, we compare in Figure~\ref{fig:purescattering_first} $\Delta a/a_{0}$ only for the first encounter. We distinguish the first encounter from the subsequent one in the resonant pure scattering case such that the semi-major axis of the binary becomes nearly constant between encounters. If two encounters are too close in time, the inflection point in $a$ (where the second derivative is equal to zero)
between the two encounters will be used as the post-scattered value of the first encounter. $\Delta a/a_{0}$ clearly becomes more aligned along the grey diagonal line, indicating that the impact of the first encounter by TDEs and by pure scatterings leads to a similar change in the orbital parameters.

On the other hand, as shown in the right panel of Figure~\ref{fig:purescattering}, the fractional changes in the eccentricity due to TDEs and pure scatterings are similar even when measured at the end of simulations. 

These findings have interesting implications. First, 
contrary to the outcome of binaries predicted in the stellar dynamics adopting the point-particle approximation, a hard binary experiencing a TDE can be either softened or hardened. This implies that in dense environments with a binary fraction $\lesssim$10\% \citep{Ivanova2005,Cool2002}, where  binary-single scatterings  are the dominant source for cluster heating and binary hardening, if the disruption of stars from close encounters is not considered in the rate estimation (typically the resonance scatterings), the hardening rate of the hard (soft) binary black holes can be overestimated (underestimated). Second, the fact that the final binary outcome of TDEs and pure scatterings could be very qualitatively different suggests that it is important to take the impact of TDEs on the binary orbit into account properly in few-body or $N-$body simulations where only pure scatterings are typically considered. In particular, one would have to extract the orbital change only after the encounter that would have led to a TDE.

\subsection{Observational signatures}

\subsubsection{Electromagnetic emission}

Unlike isolated BBHs, surrounding stellar debris produced in TDEs by BBHs can generate electromagnetic (EM) radiation. Furthermore, if the debris remains until the binary merges, there may be an electromagnetic counterpart to the gravitational wave emission from the BBH merger. Therefore, it is of great interest to find out the expected EM signatures from the interaction of the BBH  with the stellar debris. Because the resolution of our SPH simulations is not fine enough to resolve the photosphere, we cannot estimate the luminosity coming from the photosphere of the system. Therefore, we rather took an order-of-magnitude approach  to estimate the luminosity. We split a sufficiently large volume that covers most of the gas and binary ($|x|, |y| \leq 18a$ and $|z|\leq 2 a$) into $100\times100\times100$ rectangular grids. Then we calculated the cooling rate of each volume element in the $z$ direction as the ratio of the radiation energy integrated over the column to the local cooling time $\tau h c (1+u_{\rm gas}/u_{\rm rad})$ where $\tau$ is the optical depth to the mid-plane, $h$ is the density-weighted scale-height, $u_{\rm gas}/u_{\rm rad}$ is the total gas internal energy to the total radiation energy of the column. We find that the maximal total cooling rate (i.e. the sum of the per-column cooling rate) of $\simeq 10^{41}-10^{44}$ erg/s, which is super-Eddington by two to five orders of magnitude, appears almost right after the first encounter. In most of the models, the cooling rate drops to $10^{40}-10^{41}$ erg/s in less than a few hours, and this luminosity level is then maintained for more than tens of hours. Because of such a high luminosity, the radiation pressure is very large, resulting in a geometrically thick stream around the binary ($h/r\gtrsim 0.1 $). The gas around the binary is very optically thick as well (optical depth to the mid-plane $\gtrsim 10^{6}$). This allows us to assume that the radiation is fully thermalized. Assuming blackbody radiation, we estimate the luminosity-weighted average of the thermodynamic temperature. The temperature follows a similar evolution in time as the cooling rate. The blackbody temperature is $10^{5}-10^{6}$K at peak, corresponding to extreme-UV or X-rays, then drops to $10^{4}-10^{5}$K in a few hours lasting over tens of hours.

The main energy source of the luminosity we estimated above is the thermal energy dissipated by shocks created due to stream-stream and stream-BH interactions. However, the main energy component is likely to come from the release of binding energy as the debris accrete onto the BH and convert some of that energy into radiation.
 In our simulations, which have an accretion radius of $\simeq 200r_{\rm g}$ for computational reasons, the exact amount of accreted particles cannot be accurately computed. 
 Additionally, highly super-Eddington accretion likely leads to strong outflows/winds (e.g. \citealt{Skadowski2014}),  which would regulate the subsequent accretion. Furthermore, the very high accretion rate and the induced BH spin, along with magnetic fields of debris inherited from the star, can be large enough to launch a jet (e.g. \citealt{Krolik2012}). The additional luminosity powered by the jet is likely to track the accretion rate (i.e., luminosity $\propto \dot{M}c^{2}$ with an efficiency factor) and hence show a modulation in the emitted radiation, such as in gamma-ray bursts \citep{GRB}. If both black holes manage to launch jets which happen to fall within the field of view, the observed lightcurves may show higher-frequency modulations as the luminosity of each jet is superposed.

 In addition to the prompt electromagnetic signatures discussed above, the fact that a TDE may create an accretion disk around one of the BHs, or around the entire BBH depending on the TDE conditions and binary size, may lead to an additional electromagnetic counterpart at the time of merger. Indeed, if matter cools quickly enough that the magneto-rotational instability is shut down before the entire disk is accreted, as in the model proposed by \citet{Perna2016} (in that case the matter was provided by the fallback matter after the second supernova explosion), then an electromagnetic counterpart may be produced at the time of the BBH merger, as the 'dead' disk is revived at the time of the merger and accretes rapidly then. This counterpart may rather resemble a weak short gamma-ray burst.

\subsubsection{Gravitational wave emission}

Binary black hole mergers provide the most significant contribution to the
gravitational wave sources detected by the LIGO/Virgo observatory.
Our simulations suggest that  if a merging BBH experiences a TDE, the semimajor axis changes by up to $20\%$. Initially circular binaries become more eccentric and initially eccentric binaries become more eccentric or circular depending on the orientation of the incoming orbit of the star. These changes in the orbit parameters naturally lead to changes of the GW-driven merger time scale $t_{\rm GW}$. We show the fractional change $\Delta t_{\rm GW}/t_{\rm GW,0}$ relative to that for the original binary for all our models in Figure~\ref{fig:frac_tgw}. Because the eccentricity increases by a similar amount in the majority of our models (see Figure~\ref{fig:deltae_deltaa}), $\Delta t_{\rm GW}/t_{\rm GW,0}$ almost monotonically increases with $\Delta a/a_{0}$. More specifically, a TDE in a prograde (retrograde) encounter increases (decreases) $t_{\rm GW}$ by a factor up to order unity. Hence a single TDE would not alter the time delay distribution of BBH mergers significantly\footnote{Remind that our BBHs that experienced a TDE remain still in the GW regime.}. However, if binaries undergo TDEs frequently in their lifetimes, the time delay distribution and the $\chi$ distribution can be shifted depending on how frequent prograde or retrograde encounters are. Our result that the fractional changes in $a$ and $e$ due to prograde TDEs tend to be greater than those due to retrograde TDEs may suggest that the time delay distribution is shifted preferentially towards a longer time and the $\chi-$ distribution towards positively large $\chi$ even for isotropic encounters. In addition, we find that for our parameters the impact of in-plane encounters is generally greater than that of off-plane encounters (e.g., $i60$ and $i120$). Assuming that this is true for a wide range of parameters, a more dramatic alteration of the time delay distribution may be expected for binaries embedded in AGN disks. However, we have to stress that to confirm these trends, a more systematic study exploring a wider parameter space should be performed. 

\subsection{Varieties of stellar-mass BBH-driven TDEs}

In this paper, we considered close encounters between a $1\Msol$ middle-age main-sequence star and a merging BBH. The rate of such events is highly uncertain. The rate per BBH $d\mathcal{R}/dN$ may be calculated, still on an order-of-magnitude level, as `$n_{\rm s} \Sigma v$' where $n_{\rm s}$ is the number density of single stars, $v$ is the relative velocity between the single star and the BBH, $\Sigma$ is the encounter cross-section. For TDEs of $1\Msol$ stars by equal-mass BBHs, the `$n_{\rm s} \sigma v$' estimate gives  $d\mathcal{R}/dN\simeq 10^{-10}-10^{-9}\yr^{-1}$ \citep{Samsing2019, Lopez2019}. To calculate the total rate $\mathcal{R}$, one needs to integrate the per-BBH rate with star formation history, merger delay time and metallicity-dependent BBH formation efficiency, some of which are highly model-dependent. Nonetheless, the per-galaxy rate may be estimated as  $\mathcal{R}\simeq N_{\rm tot}\times (d\mathcal{R}/dN)$ (assuming $d\mathcal{R}/dN$ does not change over time) where $N_{\rm tot}$ is the number of merging BBHs per galaxy.

Although we consider encounters involving a $1\Msol$ star as a representative case, it is possible that stars with different masses can be disrupted in encounters with a BBH. In fact, if stars in clusters follow a typical stellar mass function (e.g., \citealt{Kroupa2002}), because of more abundant low-mass stars and shorter life times of more massive stars, encounters between a low-mass star (say $\lesssim 1\Msol$) and a BBH would be high exclusively in old clusters. In young clusters, mass segregation would enhance the number density of more massive stars near the potential minimum in the cluster where BHs are likely to reside, which, in turn, would increase the rate of the massive star-BBH encounters. However, unless the enhancement in the number density is substantially high, it is likely that the number of encounters involving low-mass stars would still be dominant. Because the total number of such encounters involving a single star with its life time $t_{\rm life}$ may be expressed as $\mathcal{N}\simeq \mathcal{R}t_{\rm life}$, the relative number of encounters involving a star with mass $M_{1}$ and those involving a star with mass $M_{2}$ is $\mathcal{N}_{1}/\mathcal{N}_{2}\simeq [n_{\rm s,1}\sigma_{1}t_{\rm life,1}]/[n_{\rm s,2}\sigma_{2}t_{\rm life,2}]$. Here, $\sigma$ is the velocity dispersion. Assuming the Kroupa stellar mass function following $\propto M^{-2.3}$ for $M>0.5\Msol$, we derive $n_{\rm s,1}/n_{\rm s, 2}\simeq (M_{1}/M_{2})^{-2.3}$. Using the mass-luminosity relation for main-sequence stars, which yields $t_{\rm life}\propto M^{-3}$, and assuming $\sigma_{1}\simeq \sigma_{2}$, we find $\mathcal{N}_{1}/\mathcal{N}_{2}\simeq (M_{1}/M_{2})^{-5.3}$, indicating that the number of events has a strong dependence on the stellar mass. We stress that this is a rough estimate without considering potentially important physical effects such as mass segregation.

Despite the overall small number of encounters involving high-mass stars, the individual events with high-mass stars are easier to  detect because they are brighter. Furthermore, the impact on the binary orbit and mass accretion would be greater than for encounters with low-mass stars. In this regard, encounters of high-mass stars could have a significant impact on the growth of BHs in young stellar environments \citep[e.g.,][]{Giersz+2005} and in metal-poor environments where massive stars are abundant (e.g., Population III stars in the early Universe).

The nearly parabolic obits assumed in our simulations are reasonable choices considering thar the typical eccentricity of two-body encounters in clusters with velocity dispersion $\sigma$ is $|1-e|\simeq 10^{-4}(\sigma/15\km{\rm s}^{-1})^{2}(\Mbh/40\Msol)^{-2/3}(M_{\star}/1\Msol)^{-1/3}(R_{\star}/1\Rsol)$ for $\Mbh\gg \mstar$ \footnote{The eccentricity is calculated using the relative specific kinetic energy $\simeq 0.5\sigma^{2}$ and the specific angular momentum $\simeq \sqrt{2 G\Mbh r_{\rm t}}$ for $\Mbh\gg \mstar$.}. However, it is possible that TDEs can occur in a triple or very eccentric flybys. For such cases, a larger fraction of the stellar mass is likely to remain bound to the binary than for hyperbolic or parabolic encounters. This implies that the accreted mass is likely to be higher and the events may be brighter.

Lastly, we considered one of several ways in which TDEs can take place during three-body interactions. Other possible cases include encounters between a star-BH binary and a star, a star-BH binary and a BH. In principle, as long as at least one black hole and one star are involved in three-body encounters, a TDE can happen. Importantly, the rate, the evolution of stellar debris, and thus the observational signatures, are possibly very different from one another. We will investigate the outcome of such encounters of different types in our future work.

\section{Summary and Conclusions}\label{sec:summary}

In this work we have investigated the outcome of tidal disruption events of $1\Msol$ middle-age main-sequence stars, whose initial profile was a realistic stellar model (computed with MESA), 
by LIGO/Virgo stellar-mass binary black holes. We have performed a suite of hydrodynamics simulations with a wide range of key parameters, including the semimajor axis, binary mass ratio, binary eccentricity, impact parameter and inclination.  We especially focus on studying the accretion rate and the change in the orbital parameters and spins due to TDEs. We summarize the main results as follows:
\begin{itemize}
    \item TDEs in our simulations can be categorized into two classes, depending on the morphology of the debris stream and the immediate impact of TDEs on the binary orbit: \textit{regular debris flow} and \textit{non-regular debris flow}. 
    
    \item In the \textit{regular debris flow} case, the star is disrupted outside of the binary and the shape and trajectory of the debris look very much like regular TDEs by single black holes. This case shows a few characteristic features such as a relatively mild impact on the binary orbit, symmetric mass accretion and quasi-periodic modulations of the accretion rate due to repeated perturbations of the stream by the binary. All the models where the star loses its mass partially at the first encounter (e.g., Models $\alpha1-pro.$ and $\alpha1-retro.$) and most of the retrograde encounters fall into this category. 
    
    \item In the \textit{non-regular debris flow} case, the star is disrupted by the disruptor BH with a relatively large impact parameter. Characteristic features of this case are a large momentum kick, resulting in a widening (hardening) of the binary in prograde (retrograde) encounters, asymmetric accretion and significant disruption of the debris stream by the bystander BH, followed by irregular and violent interactions between the debris and binary. For this case, the variability of the accretion rate is less prominent than it is for the \textit{regular debris flow} case. 
    
    \item The semimajor axis of the binary experiencing a prograde (retrograde) TDE tends to increase by up to $\simeq 20\%$ ($10\%$). For a wider binary ($a\geq0.03\AU$), the fractional change in $a$ is found to be larger ($\simeq 20\%$) even for retrograde TDEs. 
    
    \item The initially circular binary becomes eccentric by $\lesssim 10\%$ for both prograde and retrograde TDEs. The eccentricity of the initially eccentric binaries increases (decreases) for retrograde (prograde) TDEs. 
    
    \item The spin induced by the mass accretion is found to be small for the parameters considered in our paper ($M_{\star}/\Mbhp\simeq 0.05$). Prograde TDEs result in a change in the the effective spin parameter $\chi$ by $\lesssim 0.02$, while $\chi\gtrsim -0.005$ for retrograde TDEs. The absolute magnitude of the effective spin parameter $\chi$ has a positive correlation with the total accreted mass. 
    
    \item The immediate impact of TDEs on the binary orbit is different from that of pure scatterings especially when the interaction between the binary and the star is in the resonant scattering regime. A TDE and pure scattering change the binary orbit parameters by a similar amount at the first encounter. However the difference arises in that there would not be a subsequent scattering for TDEs, whereas the binary can undergo multiple encounters with the star for the pure scattering case, which tend to make hard binaries harder. This indicates that the hardening rate of hard (soft) binaries in dense environments could be overestimated (underestimated). 
    
    \item Based on the approximate estimate of the cooling rate using the local cooling time, TDEs by BBHs can promptly generate radiation in extreme UV to soft X-rays at the luminosity of $10^{40}-10^{44}$ erg/s over more than tens of hours. However, it is necessary to investigate the long-term evolution of the debris with proper radiation transport calculations in order to extract an  accurate lightcurve. 
    
    \item For LIGO/Virgo binaries, a single TDE can result in changes of the GW-driven merger time scale by order unity. More specifically, a TDE in a prograde (retrograde) encounter tends to increase (decrease) the merger time. Although a single TDE would not alter the time delay distribution of BBH mergers significantly, if TDEs by BBHs are frequent, the time delay distribution of BH mergers and the $\chi$ distribution can be shifted depending on how frequent prograde or retrograde encounters are. For clusters where encounters are isotropic, the time delay distribution of BH mergers may preferentially be shifted towards a longer time and and the $\chi$ distribution towards positively large $\chi$. Given that the impact of in-plane encounters is generally greater than that of off-plane encounters, there may be a greater alteration of those distributions for binaries embedded in AGN disks.
    
\end{itemize}

Tidal disruption events by stellar-mass binary black holes are one of the few events that might lead to an EM counterpart to BBH mergers\footnote{Another example would be BBH mergers in AGN disks \citep{Bartos2017,Graham2020}.}. Therefore, observation of EM counterparts of a merger event in a cluster or a galactic center of a dormant galaxy can serve as strong evidence of a TDE by a BBH before its merger. 

It is a rich physics problem that has potentially many astrophysical and observational implications, including the evolution history of binaries in clusters and AGN disks, mergers of black holes and hence GW emission. For example, such events can occur across the cosmic time. Thus the rate of the events could give us information on dense stellar environments throughout the history of the Universe. We will investigate different aspects of these events, including TDEs by BBHs and BH-neutron star binaries with magnetic field, and those in stellar-mass triples, with a focus on their EM and GW observational signatures in the future.

\section*{Acknowledgements}
The authors thank the referee for constructive feedback and suggestions, which helped us to improve the manuscript. TR is grateful to Selma de Mink, Deepika Bollimpalli and Martyna Chruslinska for useful discussions and suggestions.
This research project was conducted using computational resources (and/or scientific computating services) at the Max-Planck Computing \& Data Facility. RP and YW acknowledge support by NSF award AST-2006839.

\bibliographystyle{mnras}
%\bibliography{biblio.bib} 

\end{document}